\documentclass[aip,jcp,reprint,amsmath,amssymb,floatfix,citeautoscript]{revtex4-1}
\usepackage{cancel}
\usepackage{amsmath}
\usepackage{physics}
\usepackage{graphicx}
\usepackage{amssymb}
\usepackage{amsthm}
\usepackage{bm}
\usepackage{dcolumn}
\usepackage{braket}
\usepackage{longtable}
\usepackage{ragged2e}
\usepackage{txfonts}
\usepackage[version=3]{mhchem} 
\usepackage[T1]{fontenc}       
\usepackage{pstricks}
\usepackage{algorithm}         
\usepackage{algpseudocode}     
\usepackage{siunitx}
\usepackage[colorlinks=true,allcolors=blue]{hyperref}
\usepackage[capitalise]{cleveref}
\usepackage{multirow}

\let\vr\undefined
\newcommand{\vr}{{\bm{r}}}

\begin{document}

\title{
A simplified GW/BSE approach for charged and neutral excitation energies of large molecules and
nanomaterials
}

\author{Yeongsu Cho}
\author{Sylvia J. Bintrim}
\affiliation{Department of Chemistry, 
Columbia University, New York, New York 10027, USA}
\author{Timothy C. Berkelbach}
\affiliation{Department of Chemistry, 
Columbia University, New York, New York 10027, USA}
\affiliation{Center for Computational Quantum Physics, Flatiron Institute, New York, New York 10010, USA}
\email{tim.berkelbach@gmail.com}

\begin{abstract}
Inspired by Grimme's simplified Tamm-Dancoff density functional theory approach
[S. Grimme, J. Chem. Phys. \textbf{138}, 244104 (2013)], we describe a
simplified approach to excited state calculations within the GW approximation to
the self-energy and the Bethe-Salpeter equation (BSE), which we call sGW/sBSE.
The primary simplification to the electron repulsion integrals yields the same
structure as with tensor hypercontraction, such that our method has a storage
requirement that grows quadratically with system size and computational timing
that grows cubically with system size. The performance of sGW is
tested on the ionization potential of the molecules in the GW100 test set,
for which it differs from \textit{ab intio} GW calculations by only 0.2~eV.
The performance of sBSE (based on sGW input) is tested on the excitation energies of molecules in the 
Thiel set, for which it differs from \textit{ab intio} GW/BSE calculations by about 0.5~eV.
As examples of the systems that can be routinely studied with sGW/sBSE, we calculate
the band gap and excitation energy of hydrogen-passivated silicon nanocrystals with
up to 2650~electrons in 4678 spatial orbitals and the absorption spectra of two large organic dye molecules
with hundreds of atoms.
\end{abstract}

\maketitle

\section{Introduction}

The GW approximation to the self-energy and the Bethe-Salpeter equation (BSE)
are known to provide accurate charged and neutral excitation energies,
respectively~\cite{Hedin1965,Strinati1980,Hanke1980,Strinati1982,Strinati1984,Hybertsen1985,Hybertsen1986,Albrecht1998,Rohlfing2000}. 
Given their successful application to solid-state materials, they have been
increasingly applied to problems in molecular chemistry, for which they have
been found to be affordable approaches with reasonable
accuracy~\cite{Tiago2005,Faber2014,Koerbel2014,Bruneval2015,Jacquemin2015,Setten2015,Caruso2016,Knight2016,Rangel2017}.
We refer to the reviews presented in Refs.~\onlinecite{Blase2018,Golze2019,Blase2020} for further information.

The computational cost of \textit{ab intio} GW/BSE calculations depends on implementation details, which yield
computational timings that scale as $N^3$ to $N^6$ with system size
$N$~\cite{Foerster2011,Deslippe2012,setten2013,Govoni2015,Ljungberg2015,Krause2016,Bruneval2016,Wilhelm2018,Koval2019,Liu2020,Zhu2021,Bintrim2021}
(throughout this work we exclusively consider the non-self-consistent G$_0$W$_0$ approximation
but will typically refer to it as the GW approximation, for brevity). 
Although promising, the storage requirements and computational timing for \textit{ab intio}
GW/BSE calculations are still prohibitive for applications to very large systems
or to problems requiring many calculations, such as averaging over the course of a 
molecular dynamics trajectory or in workflows for materials screening.

The same observations about computational costs of \textit{ab intio} density functional theory (DFT)
and time-dependent DFT (TDDFT) 
have led to a number of more affordable semiempirical approximations,
including density functional tight-binding (DFTB)~\cite{Elstner1998,hourahine2020dftb+},
extended tight-binding (xTB)~\cite{Grimme2017,bannwarth2019gfn2,Bannwarth2020}.
the simplified Tamm-Dancoff approximation (sTDA) to TDDFT, and combinations thereof,
such as TD-DFTB~\cite{niehaus2001tight,Trani2011} and sTDA-xTB~\cite{grimme2016ultra}. 
Inspired in particular by Grimme's sTDA, here we present a simplified GW/BSE approach
that we call sGW/sBSE. As the heart of both approaches is a particular approximation to the
electron repulsion integrals, which we show leads to a sGW/sBSE implementation with
storage requirements that are quadratic in system size and execution times that
are cubic in system size.
Although sGW/sBSE results differ from \textit{ab intio} ones by 0.1-1~eV, they can be applied to
very large systems using only commodity computing resources.

\section{Theory}


\subsection{Integral approximations}
\label{ssec:theory_int}

A standard DFT calculation is first performed to yield the Kohn-Sham eigenvalues $\varepsilon_p$
and molecular orbitals (MOs) $\psi_p(\vr)$. The MOs are expanded in a basis of atomic orbitals (AOs)
$\phi_\mu(\vr)$ or symetrically orthogonalized AOs $\phi^\prime_\mu(\vr)$, 
\begin{equation}
    \psi_p(\vr) = \sum_\mu \phi_\mu(\vr) C_{\mu p} = \sum_\mu \phi^\prime_\mu(\vr) C^\prime_{\mu p},
\end{equation}
where $\mathbf{C}^\prime = \mathbf{S}^{1/2} \mathbf{C}$ and $\mathbf{S}$ is the AO overlap matrix. 
The primary simplification we make is to the four-center two-electron repulsion
integrals (ERIs), whose storage and manipulation is responsible for much of
the cost in correlated calculations with atom-centered basis functions.
Following Grimme~\cite{grimme2013simplified}, the MO ERIs are approximated to be (in 1122 notation)
\begin{equation}
\label{eq:eri}
(pq|rs) \approx \sum_{\mu\nu} L_{pq}^\mu L_{rs}^\nu J_{\mu\nu}
    = \sum_{\mu\nu} C^\prime_{\mu p} C^\prime_{\mu q} C^\prime_{\nu r} C^\prime_{\nu s} J_{\mu\nu}
\end{equation}
where $L_{pq}^{\mu} = C^\prime_{\mu p} C^\prime_{\mu q}$ is the orthogonalized
AO component of the orbital pair density. 
In the above, we are retaining (and approximating) only one- and two-center
Coulomb integrals 
in the orthogonalized AO basis.
The one-center Coulomb integrals are evaluated exactly,
$J_{\mu\nu}^{(\mathrm{1c})} = (\mu\mu|\nu\nu)$, and the two-center Coulomb
integrals are approximated by the Mataga-Nishimoto-Ohno-Klopman
formula~\cite{nishimoto1957electronic,ohno1964some,klopman1964semiempirical},
\begin{equation}
    J_{\mu\nu}^{(\mathrm{2c})} \approx \left(\frac{1}{|r_\mu-r_\nu|^2+\eta_{\mu\nu}^{-2}}\right)^{-1/2},
\end{equation}
where
$\eta_{\mu\nu} = \left[(\mu\mu|\mu\mu)+(\nu\nu|\nu\nu)\right]/2$.
The use of exact one-center integrals is a slight departure from the original
sTDA method~\cite{grimme2013simplified}.

The ERI approximation~(\ref{eq:eri}) has the same structure as the density
fitting approximation (also known as the resolution of the identity
approximation)~\cite{Whitten1973,Vahtras1993,Feyereisen1993,Werner2003}, 
which have been regularly used in \textit{ab intio}
GW/BSE implementations with atom-centered basis sets~\cite{Ren2012,Wilhelm2016,Krause2016,Wilhelm2018,Koval2019,Liu2020,Zhu2021}. 
However, we emphasize that the three-index tensors
$L_{pq}^\mu$ are expressible as a product of two-index objects, the MO coefficients.
Therefore, the standard density fitting procedure, requiring the calculation of two-center
and three-center integrals and the solution of a system of linear equations, is completely
bypassed (although we note that it may be used in the initial DFT calculation).
More importantly, the final equality of Eq.~(\ref{eq:eri}) has the same structure as
the ERIs with tensor hypercontraction~\cite{Hohenstein2012}, leading to an 
sGW/sBSE implementation that only requires the storage of two-index objects and
opportunities for reductions in the scaling of the computational time.

\subsection{Simplified GW}

For the remainder of the manuscript, we will use $i,j,k,l$ to index occupied MOs
in the Kohn-Sham reference, $a,b,c,d$ to index unoccupied MOs, and $p,q,r,s$ for general MOs.
Within the diagonal G$_0$W$_0$ approximation, matrix elements of the self-energy operator are given by
\begin{equation}
\label{eq:sigma}
\Sigma_{p}(\omega) = \Sigma_{p}^{\mathrm{(c)}}(\omega) - \sum_i (pi|ip).
\end{equation}
The second term is the bare exchange part and the first term is the correlation part
\begin{equation}
\label{eq:Sigma_c}
\begin{split}
\Sigma_p^{\mathrm{(c)}}(\omega)
    = \frac{i}{2\pi}\sum_q\int d\omega^\prime \frac{W_{pqqp}(\omega)-(pq|qp)}
        {\omega-\omega^\prime-\varepsilon_q+i\eta\textrm{sgn}(\varepsilon_q-\mu)},
\end{split}
\end{equation}
where $\eta$ is a positive infinitesimal and $\mu$ is the chemical potential. The 
screened Coulomb interaction is
\begin{equation}
\label{eq:W}
W(\vr_1,\vr_2;\omega) = \int d\vr \varepsilon^{-1}(\vr_1,\vr;\omega) |\vr-\vr_2|^{-1}
\end{equation}
where the dielectric function is
\begin{equation}
\varepsilon(\vr_1,\vr_2;\omega) = \delta(\vr_1-\vr_2) - \int d\vr |\vr_1-\vr|^{-1} P(\vr,\vr_2;\omega)
\end{equation}
and the independent-particle polarizability is
\begin{equation}
\begin{split}
P(\vr_1,\vr_2;\omega) &= \sum_{ia}
    \psi_a(\vr_1)\psi_i(\vr_1)\psi_i(\vr_2)\psi_a(\vr_2) \\ 
    & \hspace{1em} \times
    \left[\frac{1}{\omega-(\varepsilon_a-\varepsilon_i)+i\eta}
    -\frac{1}{\omega+(\varepsilon_a-\varepsilon_i)+i\eta}\right].
\end{split}
\end{equation}

As mentioned above, Eq.~(\ref{eq:eri}) has the same structure as the density fitting approximation
with $|\phi^\prime_\mu(\vr)|^2$ playing the role of the auxiliary basis,
\begin{equation}
\psi_p(\vr)\psi_q(\vr) 
    \approx \sum_\mu C_{\mu p}^\prime C_{\mu q}^\prime |\phi_\mu^\prime(\vr)|^2
    = \sum_\mu L_{pq}^{\mu} |\phi_\mu^\prime(\vr)|^2.
\end{equation}
In this nonorthogonal auxiliary basis, the dielectric function has 
a matrix representation $\varepsilon_{\mu\nu}(\omega) \equiv [\bm{\varepsilon}(\omega)]_{\mu\nu}$ with
\begin{align}
\bm{\varepsilon}(\omega) &= \mathbf{S}^\prime - \mathbf{J}\mathbf{S}^{\prime-1} \mathbf{P}(\omega), \\
S_{\mu\nu}' &= \int d\vr |\phi^\prime_\mu(\vr)|^2 |\phi^\prime_\nu(\vr)|^2, \\
P_{\mu\nu}(\omega) &= \int d\vr_1 d\vr_2 |\phi^\prime_\mu(\vr_1)|^2 P(\vr_1,\vr_2;\omega) |\phi^\prime_\nu(\vr_2)|^2.
\end{align}
Rather than inverting the dielectric matrix at every frequency and numerically integrating, we use the
plasmon-pole approximation~\cite{Hybertsen1986,von1988precise,Larson2013}. Considering 
the generalized eigenvalue problem
\begin{equation}
\bm{\varepsilon}(\omega) \mathbf{U}(\omega) = \mathbf{S}^\prime \mathbf{U}(\omega) \bm{\lambda}(\omega),
\end{equation}
we assume $\mathbf{U}(\omega) = \mathbf{U}(\omega=0)$ and that the eigenvalues can be
parameterized by the form
\begin{equation}
\lambda_l^{-1}(\omega) = 1 + z_l\left(\frac{1}{\omega-(\omega_l-i\eta)}-\frac{1}{\omega+(\omega_l-i\eta)}\right).
\end{equation}
The parameters $z_l$ and $\omega_l$ are chosen to match the numerical eigenvalues $\lambda_l(\omega)$ obtained at the two
frequencies $\omega = 0$ and $\omega = \varepsilon_\mathrm{gap}$ (i.e., the Kohn-Sham band gap).
With this form, the frequency integration can be performed analytically to give
\begin{equation}
\label{eq:sigma_ppa}
\begin{split}
\Sigma^{\mathrm{(c)}}_p(\omega)
    &= \sum_{q\mu\nu} L_{pq}^\mu
        [\mathbf{S}^\prime\mathbf{U}\bm{\Lambda}^{(q)}(\omega)
            \mathbf{U}^{-1}\mathbf{S}^{\prime-1}\mathbf{J}]_{\mu\nu}L_{qp}^\nu \\
    &= \sum_{q l} \left[ \sum_\mu C_{\mu p}' C_{\mu q}' [\mathbf{S}^\prime\mathbf{U}]_{\mu l} \right]
        \Lambda^{(q)}_{l l}(\omega) \\
    &\hspace{2em} \times \left[ \sum_\nu [\mathbf{U}^{-1}\mathbf{S}^{\prime-1}\mathbf{J}]_{l \nu}
            C_{\nu p}' C_{\nu q}'\right]
\end{split}
\end{equation}
where $\bm{\Lambda}^{(q)}(\omega)$ is a diagonal matrix with elements 
\begin{equation}
    \Lambda_{ll}^{(q)}(\omega)=\frac{z_l}{\omega-\varepsilon_q-\omega_l\textrm{sgn}(\varepsilon_q-\mu)}.
\end{equation}
We calculate the GW quasiparticle energies $E_p$ using the linearized form
\begin{equation}
E_p = \varepsilon_p + Z_p \left[ \Sigma_p(\varepsilon_p) - v^{\mathrm{(xc)}}_{pp} \right]
\end{equation}
where $v^{\mathrm{(xc)}}_{pp}$ is a diagonal matrix element of the DFT exchange-correlation potential
and the renormalization factor is
\begin{equation}
Z_p = \left(1-\frac{\partial \Sigma(\omega)}{\partial \omega}\Big|_{\varepsilon_p}\right)^{-1}.
\end{equation}
Using only $O(N^2)$ storage, the intermediates indicated in square brackets in Eq.~(\ref{eq:sigma_ppa}) can be formed for
each orbital $p$ of interest in $O(N^3)$ time; the time needed to calculate $n_\mathrm{GW}$ eigenvalues is then
$O(n_\mathrm{GW}N^3)$. If $O(N^3)$ storage is available, then the intermediates indicated
can be calculated once and stored as a three-index object; the time needed to calculate \textit{all} eigenvalues
is then only $O(N^3)$.

Finally, we note that the full self-energy~(\ref{eq:sigma}) requires bare MO exchange integrals. Within
sGW/sBSE, three options exist. When the initial DFT calculation is done with a pure local functional, then
the MO exchange integrals can be approximated by Eq.~(\ref{eq:eri}). This allows the exchange contribution
for \textit{all} matrix elements of the self-energy to be calculated in $O(N^3)$ time.
In numerical tests (not shown), this was found to be a poor approximation, which we believe
to be a worthy topic of future study. Therefore, as a second option, we consider
modifying the ERI approximation to include one-center AO exchange integrals,
\begin{equation}
\label{eq:eri_1cK}
(pq|rs) \approx \sum_{\mu\nu} L_{pq}^\mu L_{rs}^\nu J_{\mu\nu}
    + \sum_{\mu\neq\nu}^{(\mathrm{1c})} \left( L_{pr}^\mu L_{qs}^\nu 
    + L_{ps}^\mu L_{qr}^\nu\right) K_{\mu\nu}^{(\mathrm{1c})}.
\end{equation}
The one-center AO exchange integrals are empirically scaled by a single
parameter, $K_{\mu\nu} = \alpha_K (\mu\nu|\mu\nu)$, and the AO ERIs
$(\mu\nu|\mu\nu)$ are calculated exactly. This improvement adds negligible computational cost
with only $O(N^2)$ scaling.
To motivate the third and final option, we recall that hybrid functionals often provide a better starting
point for G$_0$W$_0$/BSE calculations~\cite{Caruso2016,Jacquemin2017,Blase2020}.
In this case, matrix elements of the exchange operator are already available and
can be reused for free in the evaluation of the GW self-energy. We will present
results for both the second and third options in Sec.~\ref{ssec:results_sGW}.

\subsection{Simplified BSE}

Within the Tamm-Dancoff and static screening approximations, the BSE is an eigenvalue problem for the matrix
\begin{equation}
    A_{ia,jb}=(E_a-E_i)\delta_{ij}\delta_{ab}
    +\alpha(ia|jb)-(ij|W|ab),
\end{equation}
where $\alpha=2$ for singlets and 0 for triplets. 
Note that GW quasiparticle energies $E_p$ are required as input to a BSE calculation; sGW
energies will be used in sBSE calculations.
With the integral simplification~(\ref{eq:eri}) and the static screening approximation to the
screened Coulomb interaction~(\ref{eq:W}), in sBSE we have 
\begin{subequations}
\label{eq:W_eri}
\begin{align}
    (ij|W|ab) &= \sum_{\mu\nu}L_{ij}^\mu L_{ab}^\nu W_{\mu\nu} 
    = \sum_{\mu\nu} C^\prime_{\mu i} C^\prime_{\mu j} C^\prime_{\nu a} C^\prime_{\nu b} W_{\mu\nu} \\
    \mathbf{W} &=
        \mathbf{S}^\prime\bm{\varepsilon}^{-1}(\omega=0)\mathbf{J}
\end{align}
\end{subequations}
Note that $(ij|W|ab)$ has the same structure as the bare $(ij|ab)$, with
$W_{\mu\nu}$ replacing $J_{\mu\nu}$, and that $W_{\mu\nu}$ can be built
simply by matrix multiplication, requiring only quadratic storage and cubic CPU time.
We discuss further computational costs of sBSE below.

For comparison, we also provide results obtained by the simplified 
TDA (sTDA) approach~\cite{grimme2013simplified},
which is a structurally identical eigenvalue problem for the matrix
\begin{equation}
    A_{ia,jb}= 
    (\varepsilon_a-\varepsilon_i)\delta_{ij}\delta_{ab}
    +\alpha(ia|jb)-a_x(ij|ab)
    \label{eq:stda},
\end{equation}
where $a_x$ is the fraction of exact exchange included in the DFT functional.
Our sTDA calculations closely follow Ref.~\onlinecite{grimme2013simplified} except
that we use exact one-center integrals, as mentioned in Sec.~\ref{ssec:theory_int}.
Clearly the only two differences between sBSE and sTDA are the use of sGW or
DFT eigenvalues and the use of a screened Coulomb interaction or a rescaled
bare Coulomb interaction.
For solids or heterogeneous nanostructures, it is expected that the screening 
in sBSE provides a more accurate treatment of the electron-hole
interaction.

Select eigenvalues of the sTDA or sBSE matrices can be found by iterative eigensolvers,
like the Davidson algorithm, that require only matrix-vector products and the $O(N^2)$ storage
of trial vectors $c_{i}^{a}$.
With the integral approximations~(\ref{eq:eri}) and (\ref{eq:W_eri}), the sTDA and sBSE 
matrix-vector product can be done with $O(N^2)$ storage in $O(N^3)$ time,
\begin{equation}
\begin{split}
[\mathbf{Ac}]_{ia} &= (E_a-E_i)c_i^a
    + \alpha\sum_{\mu} C^\prime_{\mu i} C^\prime_{\mu a} \sum_\nu J_{\mu\nu}
        \sum_{jb} C^\prime_{\nu j} C^\prime_{\nu b} c_j^b \\
    &\hspace{1em} - \sum_{\mu} C^\prime_{\mu i} \sum_\nu C^\prime_{\nu a} W_{\mu\nu}
        \sum_{j} C^\prime_{\mu j} \sum_b C^\prime_{\nu b} c_j^b
\end{split}
\end{equation}
with intermediate formation as indicated.
In fact, if only a few sBSE eigenvalues are desired and only $O(N^2)$ storage is available, 
then the calculation of the sGW 
eigenvalues to be used in the sBSE is more expensive than
finding the few eigenvalues of the sBSE matrix.

\section{Results}
\label{sec:results}

All calculations (DFT, GW/BSE, and sGW/sBSE) were performed using a locally
modified version of the PySCF software package~\cite{sun2018pyscf,Sun2020,Zhu2021}
using the TZVP basis set \cite{schafer1994fully}.
DFT calculations used the B3LYP exchange-correlation functional unless stated otherwise.

\subsection{Simplified GW}
\label{ssec:results_sGW}

The performance of sGW is assessed with the first ionization potential (highest occupied
molecular orbital energy) of the atoms and molecules in the GW100 test set~\cite{Setten2015};
the TZVP basis set is used for all atoms except I, Xe, and Rb for which the DZVP basis
set~\cite{godbout1992optimization} is used. 
We assess the accuracy of sGW by comparing to \textit{ab intio}, full-frequency GW calculations 
(i.e., without the plasmon pole approximation) using the
same basis set.

\begin{figure}[t]
    \centering
    \includegraphics[scale=0.95]{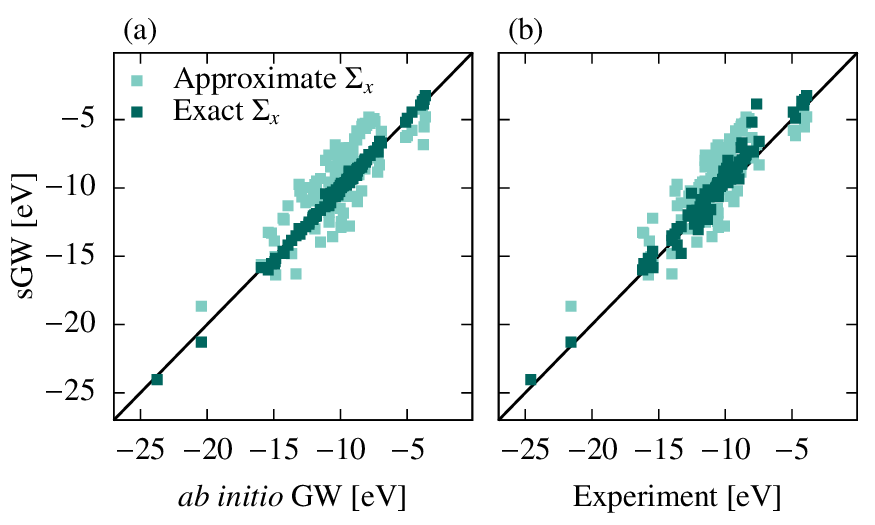}
    \caption{First ionization potential of the molecules in the GW100 set calculated by sGW@B3LYP compared 
    to (a) \textit{ab initio} GW@B3LYP results and (b) experiment, using an approximate (light) and
exact (dark) bare exchange matrix element.}
    \label{fig:gw100}
\end{figure}

We first address the evaluation of the bare exchange part of the self-energy. As mentioned
previously, it 
can be approximated by Eq.~(\ref{eq:eri}), Eq.~(\ref{eq:eri_1cK}), or calculated exactly, which is free
when a hybrid functional is used in the DFT reference. 
Figure~\ref{fig:gw100}(a) compares the IPs from \textit{ab intio} GW to those of sGW 
when approximate [Eq.~(\ref{eq:eri_1cK})] or exact exchange integrals are used.
With approximate exchange integrals, the single free parameter $\alpha_K$ is optimized to
minimize the mean absolute error (MAE) with respect to the \textit{ab intio} GW calculations, 
which leads to $\alpha_K = 0.46$; this value was found
to be robust to the basis set or exchange-correlation functional used. 
This approximate treatment of exchange integrals gives a reasonable estimate of the IP 
with a MAE of 1.81~eV (note that the IPs of the test set range from $-25$ to $-3$~eV). 
The use of the exact exchange integrals greatly increases the 
accuracy, giving a MAE of only 0.20~eV. 
Figure~\ref{fig:gw100}(b) shows that sGW also gives good agreement with
experimental IPs~\cite{Setten2015}, especially when exact exchange is used.
Therefore, for the
rest of this study, the exact exchange integrals are used, but we will return to this point
in Sec.~\ref{sec:conc}.

Table~\ref{tab:gw100} summarizes the MAE and mean signed error (MSE)
of the \textit{ab initio} GW and sGW IPs compared to experimental values.
The GW and sGW calculations used PBE, PBE0,
and B3LYP references; for PBE, we calculated the exact diagonal matrix element
of the exchange operator after the SCF convergence. 
Calculations using the hybrid functionals PBE0 and B3LYP are about 0.2~eV more
accurate than those with the PBE functional. The difference in the performance
of the \textit{ab initio} GW and sGW is marginal, indicating an accurate
estimation of the correlation term of the self-energy by sGW.
As will be demonstrated in Sec.~\ref{ssec:results_apps}, the cost of the sGW calculations
is significantly smaller than that of the \textit{ab intio} GW calculations.


\begin{table}[b]
\caption{Mean absolute error (MAE) and mean signed error (MSE) of the first ionization
potential of the molecules in the GW100 test set with respect to \textit{ab initio} GW and experimental values.
Errors are in eV and exact exchange matrix elements
were used in the sGW calculations.}
\label{tab:gw100}
\begin{ruledtabular}
\begin{tabular}{lcccccc}
     & \multicolumn{2}{c}{PBE} & \multicolumn{2}{c}{PBE0} & \multicolumn{2}{c}{B3LYP} \\
     & GW         & sGW        & GW         & sGW         & GW          & sGW         \\ \hline
MAE wrt GW & -          & 0.23       & -          & 0.19        & -           & 0.20        \\
MSE wrt GW & -          & $-0.01$      & -          & $-0.03$       & -           & $-0.04$       \\
MAE wrt Expt & 0.89       & 0.93       & 0.68       & 0.69        & 0.72        & 0.72        \\
MSE wrt Expt & 0.85       & 0.84       & 0.57       & 0.53        & 0.62        & 0.58       
\end{tabular}
\end{ruledtabular}
\end{table}

\begin{figure}[t]
    \centering
    \includegraphics[scale=0.95]{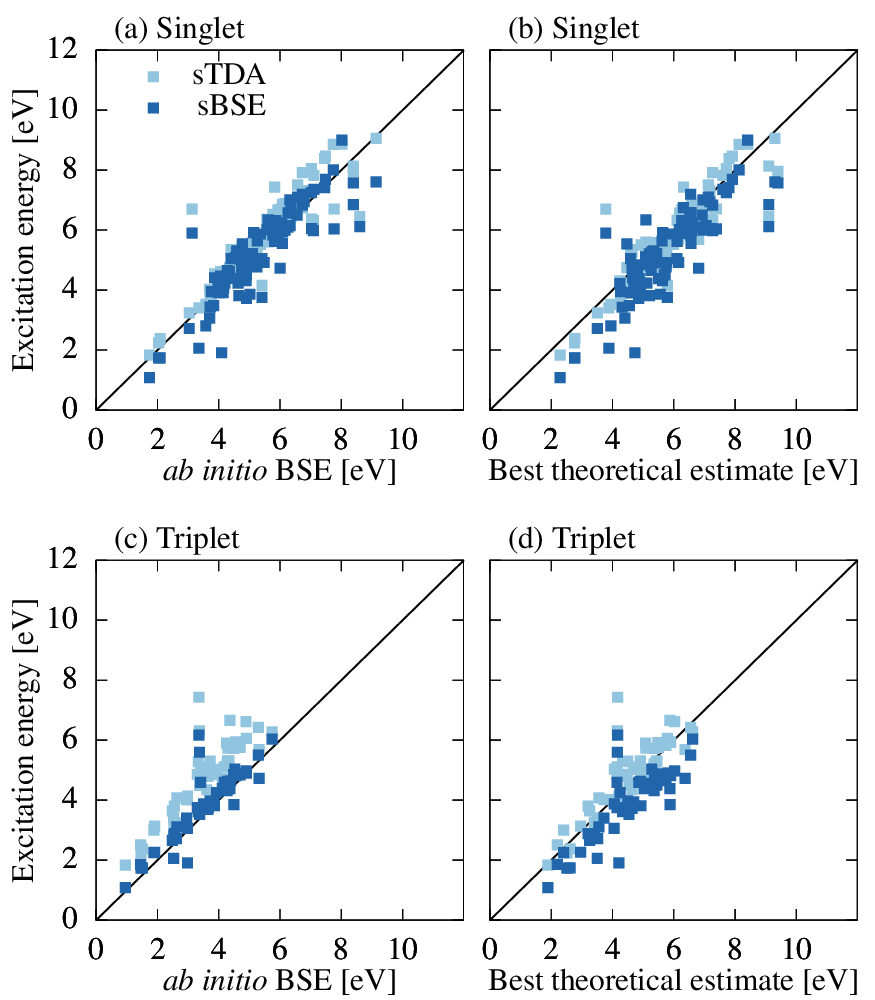}
    \caption{Singlet excitation energies of the 28 molecules in Thiel's set
calculated by sBSE compared to (a) \textit{ab initio} sBSE results and (b) best
theoretical estimtes. Analogous results for triplet excitation energies are
shown in (c) and (d).}
    \label{fig:thiel}
\end{figure}

\begin{table}[b]
\caption{Mean absolute error (MAE) and mean signed error (MSE) of singlet and triplet excitation energies of 
Thiel's set with respect to \textit{ab initio} BSE and best theoretical estimates (BTE). Errors are in eV.}
\label{tab:thiel}
\begin{ruledtabular}
\begin{tabular}{lcccccc}
     & \multicolumn{3}{c}{Singlet} & \multicolumn{3}{c}{Triplet} \\
     & BSE     & sBSE    & sTDA    & BSE      & sBSE    & sTDA   \\ \hline
MAE wrt BSE & -       & 0.51    & 0.47    & -        & 0.38    & 1.22   \\
MSE wrt BSE & -       & $-0.10$   & 0.31    & -        & 0.27    & 1.22   \\ 
MAE wrt BTE & 0.50    & 0.71    & 0.44    & 0.93     & 0.81    & 0.47   \\
MSE wrt BTE & $-0.46$   & $-0.56$   & $-0.15$   & $-0.93$    & -0.66   & 0.29  
\end{tabular}
\end{ruledtabular}
\end{table}

\subsection{Simplified BSE}

Neutral excitation energies calculated by sTDA and sGW/sBSE are tested on a set of 28
organic molecules commonly known as Thiel's set~\cite{schreiber2008benchmarks,silva2008benchmarks}.
Figure~\ref{fig:thiel} compares 97 singlet states and 51 triplet states
calculated by sBSE to those calculated by $\textit{ab initio}$ BSE and 
to the best theoretical estimates from higher level methods 
proposed by Thiel and coworkers~\cite{schreiber2008benchmarks,silva2008benchmarks}. 
The \textit{ab initio} BSE calculations are done using full-frequency GW eigenvalues,
with \textit{ab initio} static screening of the Coulomb interaction (i.e., without the
plasmon-pole approximation),
and without the Tamm-Dancoff approximation.
The MAEs and MSEs are summarized in Tab.~\ref{tab:thiel}.
The MAEs of singlet and triplet excitations calculated by sBSE with respect to
\textit{ab initio} BSE are 0.51~eV and 0.38~eV, respectively, which are similar
to the errors exhibited by sGW. 
The MAEs of singlet and triplet excitations calculated by sBSE with respect to
the theoretical best estimates are 0.71~eV and 0.81~eV, respectively, which are similar
to the errors exhibited by \textit{ab initio} BSE.
Once again we conclude that the performance difference between \textit{ab initio} BSE
and sBSE is marginal.
Interestingly, we note that sTDA gives similar but slightly smaller errors when
compared to the best theoretical estimates.


\subsection{Applications}
\label{ssec:results_apps}

\begin{figure}[t]
    \centering
    \includegraphics{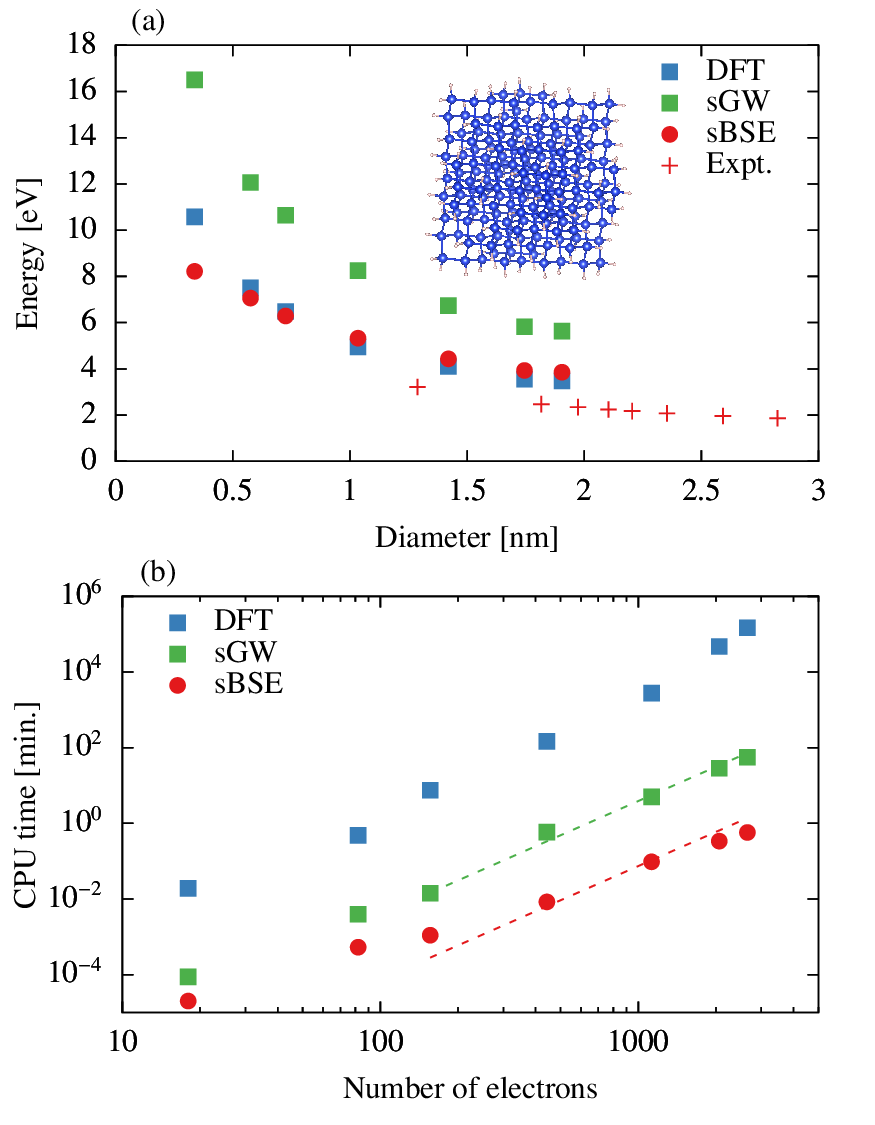}
    \caption{(a) Quasiparticle gap calculated by DFT (B3LYP) and sGW@B3LYP and first
excitation energy calculated by sBSE of silicon nanoclusters. Experimental
results (plus signs) were determined in Ref.~\onlinecite{wolkin1999electronic}
by photoluminescence. Also shown is the molecular structure of the largest studied cluster,
Si$_{181}$H$_{116}$, which has 2650 electrons and 4678 orbitals in the TZVP basis set. 
(b) CPU time required for DFT and CPU time required per eigenvalue for sGW and sBSE calculations.
The sBSE calculations used a truncated set of orbitals. The dashed lines show
$N^3$ power laws.}
    \label{fig:si}
\end{figure}

Having demonstrated the accuracy of the sGW/sBSE framework on benchmark sets of small molecules,
we move on to study silicon clusters as a prototypical semiconductor nanomaterial.
Specifically, we study hydrogen-passivated silicon clusters ranging from SiH$_4$ to
Si$_{181}$H$_{116}$, which has 2650 electrons and 4678 spatial orbitals.
Calculations on larger clusters were limited by the cost of the initial DFT
calculation.
The structure of SiH$_4$ is from the GW100 set
\cite{van2015gw}, the structures of Si$_5$H$_{12}$ and Si$_{10}$H$_{16}$ are from PubChem
\cite{pubchem}, and the structures of larger clusters are from CSIRO Nanostructure Data Bank
\cite{wilson2014shape}, without further geometry relaxation. 
In Fig.~\ref{fig:si}(a), we show the quasiparticle gap (calculated by DFT and sGW) 
and the first neutral excitation energy (calculated by sBSE) as a function of the cluster
diameter, which is estimated by approximating the cluster as a sphere with a number density
equal to that of bulk silicon (50 atoms/nm$^{3}$).
For comparison, we show experimental photoluminescence energies from 
Ref.~\onlinecite{wolkin1999electronic}. The large system sizes accessible with sGW/sBSE allow
us to compare directly to these experimental values. We see that the sBSE excitation
energies are about 1~eV higher than experiment, which may be due to a vibrational Stokes
shift, finite-temperature effects, differences in structure, or inaccuracies in the GW/BSE
level of theory.


Figure~\ref{fig:si}(b) shows the CPU time of each method as a
function of number of electrons in the silicon clusters. In practice, most calculations
are performed with some degree of parallelism using up to 32 cores; 
thus, while we report the total CPU time, the wall time can be significantly less.
For sGW, we report the CPU time required per eigenvalue using the algorithm  
that requires only $O(N^2)$ storage, such that we expect $O(N^3)$ scaling,
which is confirmed numerically. The savings afforded by our sGW algorithm enabled us to
calculate about 2500 GW orbital energies on our largest system 
(with 2650 electrons and 4678 total orbitals) in about three days
using a single 32-core node.
The sBSE calculations were performed in a truncated space that included those
orbitals with energies between 30~eV below the highest occupied orbital and 
30~eV above the lowest unoccupied orbital. Again we report the CPU time required per eigenvalue,
such that we expect $O(N^3)$ scaling, which is confirmed
numerically. In practice, we were able to calculate 50 excitation energies on the largest nanocluster
in less than an hour. We note that the sBSE calculations are less expensive than the
preceding sGW calculations.

\begin{figure}[t]
    \centering
    \includegraphics{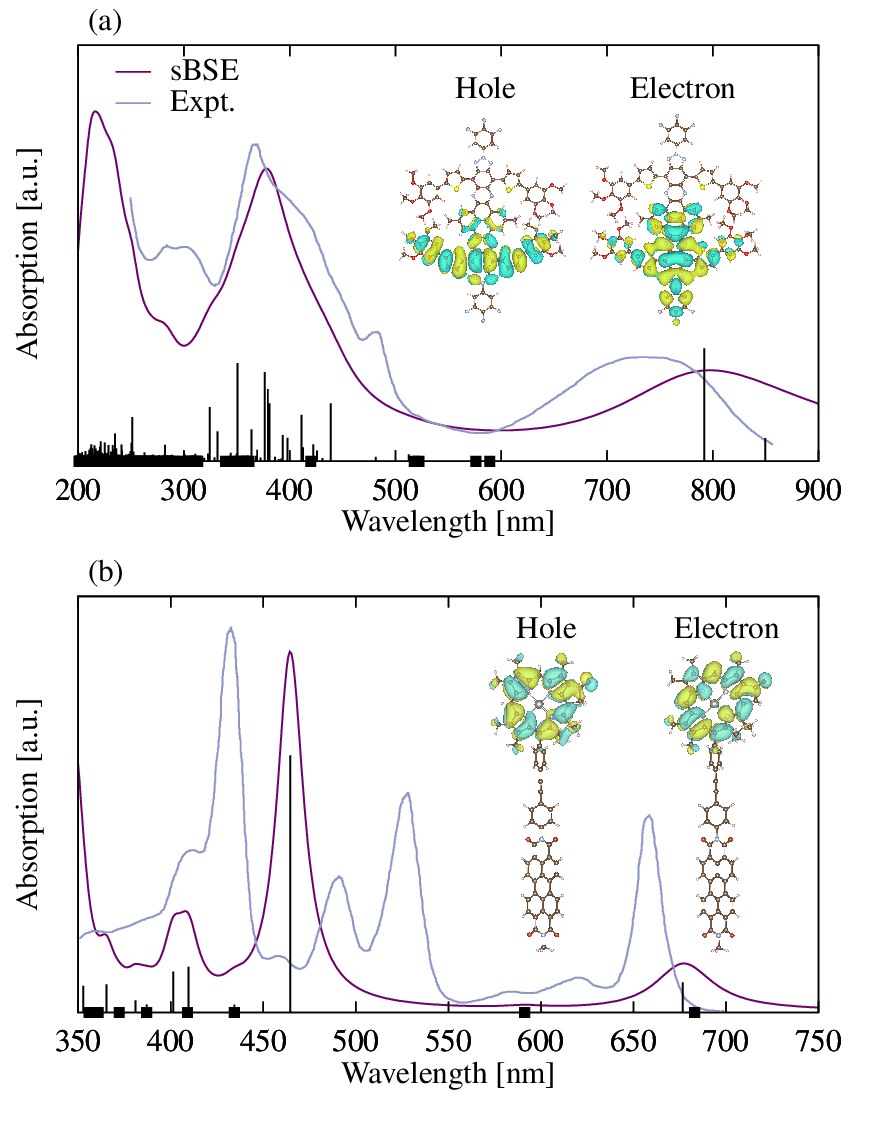}
    \caption{Absorption spectra of (a) an organic dye molecule and (b) a
chlorophyll-based donor-bridge-acceptor molecule. Black vertical lines indicate
the relative oscillator strength of each transition calculated by sBSE. Squares
show the transitions with vanishing oscillator strength. Natural transition
orbitals of the transitions at 792 nm (a) and 677 nm (b) are shown as insets.
}
    \label{fig:spec}
\end{figure}

As a final class of example problems, we apply sGW/sBSE to study the optical properties of large
organic molecules. In Fig.~\ref{fig:spec}, we show the absorption spectra
of a 192-atom organic dye molecule and a 126-atom chlorophyll-based donor-bridge-acceptor dyad,
whose structures and experimetal spectra are taken from Refs.~\onlinecite{grimme2016ultra}
and~\onlinecite{huang2016photoinduced}, respectively.
Considering that solvent effects and vibrational dynamics are neglected, the agreement
between sBSE and experiment is quite good. 
As shown in Fig.~\ref{fig:spec}(a), the organic dye molecule has a broad peak between 700
and 800~nm and strong peak at 360~nm; the
sBSE predicts a slightly redshifted broad peak, but correctly predicts the strong peak
at 360~nm, including its lineshape. As shown in Fig.~\ref{fig:spec}(b), the 
chlorophyll molecule has two prototypical strong peaks, the so-called Q band and 
Soret band. The sBSE reproduces the Q~band near 650~nm and the strong Soret band around 430 nm. 
However, we note that some transitions are spectroscopically dark at the equilibrium geometry
but the experimental spectrum shows clear vibronic signatures, indicating the likely importance
of nuclear dynamics for total agreement.
Despite being a semiempirical method, sBSE yields a proper excited-state wavefunction,
enabling a variety of analyses. As an example, in Fig.~\ref{fig:spec}, we show the largest-weight 
natural transition orbitals (NTOs)~\cite{Martin2003} of one low-lying transition
of each molecule (NTOs are the electron-hole orbital pair that best represent
the transition). Clearly, the NTOs show that the two analyzed excitations are
relatively localized to specific regions of the molecules.


\section{Conclusions and future work}
\label{sec:conc}

We have presented the simplified GW and BSE methods, which we call sGW/sBSE.
In addition to the plasmon-pole and static screening approximations, an approximation
to the electron repulsion integrals first used in Ref.~\onlinecite{grimme2013simplified}
is most responsible for the low cost of sGW/sBSE. 
The sGW/sBSE results are in good agreement with \textit{ab initio} results as well as those
of experiments or higher-level methods. In its present form, we expect that sGW/sBSE
can facilitate rapid, semiquantitative calculations of charged and neutral excitations
of large molecules and nanomaterials.

An obvious limitation of the present sGW/sBSE framework is its reliance on an
initial \textit{ab initio} DFT calculation, especially when hybrid functionals
are used, as can be seen from Fig.~\ref{fig:si}(b). Future work will address the
replacement of DFT with a semiempirical mean-field theory, similar to the
combination of the extended tight-binding method (xTB) with the sTDA for
extremely affordable calculations of excitation energies~\cite{grimme2016ultra}.
Additionally, we plan to implement spin-orbit coupling and Brillouin zone
sampling for periodic systems.  More generally, sGW/sBSE could be made into a
more \textit{ab initio} method by pursuing similar structure through tensor
decompositions or integral screening. Lastly, we believe that the sGW/sBSE
framework could be used as an affordable testing ground for improvements to the
GW/BSE formalism, such as self-consistency, vertex
corrections~\cite{Shishkin2007,Romaniello2009,Maggio2017,Lewis2019}, or the
combination of GW with dynamical mean-field theory~\cite{Biermann2003,Zhu2021a}.

\section*{Acknowledgements}

This work was supported in part by the National Science Foundation under Grant
No.~OAC-1931321 (Y.C.).  and by the National Science Foundation Graduate
Research Fellowship under Grant No.~DGE-1644869 (S.J.B.) 
We acknowledge computing resources from Columbia University's
Shared Research Computing Facility project, which is supported by NIH Research
Facility Improvement Grant 1G20RR030893-01, and associated funds from the New
York State Empire State Development, Division of Science Technology and
Innovation (NYSTAR) Contract C090171, both awarded April 15, 2010. The Flatiron
Institute is a division of the Simons Foundation.


\begin{thebibliography}{73}%
\makeatletter
\providecommand \@ifxundefined [1]{%
 \@ifx{#1\undefined}
}%
\providecommand \@ifnum [1]{%
 \ifnum #1\expandafter \@firstoftwo
 \else \expandafter \@secondoftwo
 \fi
}%
\providecommand \@ifx [1]{%
 \ifx #1\expandafter \@firstoftwo
 \else \expandafter \@secondoftwo
 \fi
}%
\providecommand \natexlab [1]{#1}%
\providecommand \enquote  [1]{``#1''}%
\providecommand \bibnamefont  [1]{#1}%
\providecommand \bibfnamefont [1]{#1}%
\providecommand \citenamefont [1]{#1}%
\providecommand \href@noop [0]{\@secondoftwo}%
\providecommand \href [0]{\begingroup \@sanitize@url \@href}%
\providecommand \@href[1]{\@@startlink{#1}\@@href}%
\providecommand \@@href[1]{\endgroup#1\@@endlink}%
\providecommand \@sanitize@url [0]{\catcode `\\12\catcode `\$12\catcode
  `\&12\catcode `\#12\catcode `\^12\catcode `\_12\catcode `\%12\relax}%
\providecommand \@@startlink[1]{}%
\providecommand \@@endlink[0]{}%
\providecommand \url  [0]{\begingroup\@sanitize@url \@url }%
\providecommand \@url [1]{\endgroup\@href {#1}{\urlprefix }}%
\providecommand \urlprefix  [0]{URL }%
\providecommand \Eprint [0]{\href }%
\providecommand \doibase [0]{http://dx.doi.org/}%
\providecommand \selectlanguage [0]{\@gobble}%
\providecommand \bibinfo  [0]{\@secondoftwo}%
\providecommand \bibfield  [0]{\@secondoftwo}%
\providecommand \translation [1]{[#1]}%
\providecommand \BibitemOpen [0]{}%
\providecommand \bibitemStop [0]{}%
\providecommand \bibitemNoStop [0]{.\EOS\space}%
\providecommand \EOS [0]{\spacefactor3000\relax}%
\providecommand \BibitemShut  [1]{\csname bibitem#1\endcsname}%
\let\auto@bib@innerbib\@empty
\bibitem [{\citenamefont {Hedin}(1965)}]{Hedin1965}%
  \BibitemOpen
  \bibfield  {author} {\bibinfo {author} {\bibfnamefont {L.}~\bibnamefont
  {Hedin}},\ }\href {\doibase 10.1103/physrev.139.a796} {\bibfield  {journal}
  {\bibinfo  {journal} {Phys. Rev.}\ }\textbf {\bibinfo {volume} {139}},\
  \bibinfo {pages} {A796} (\bibinfo {year} {1965})}\BibitemShut {NoStop}%
\bibitem [{\citenamefont {Strinati}, \citenamefont {Mattausch},\ and\
  \citenamefont {Hanke}(1980)}]{Strinati1980}%
  \BibitemOpen
  \bibfield  {author} {\bibinfo {author} {\bibfnamefont {G.}~\bibnamefont
  {Strinati}}, \bibinfo {author} {\bibfnamefont {H.~J.}\ \bibnamefont
  {Mattausch}}, \ and\ \bibinfo {author} {\bibfnamefont {W.}~\bibnamefont
  {Hanke}},\ }\href {\doibase 10.1103/physrevlett.45.290} {\bibfield  {journal}
  {\bibinfo  {journal} {Phys. Rev. Lett.}\ }\textbf {\bibinfo {volume} {45}},\
  \bibinfo {pages} {290} (\bibinfo {year} {1980})}\BibitemShut {NoStop}%
\bibitem [{\citenamefont {Hanke}\ and\ \citenamefont {Sham}(1980)}]{Hanke1980}%
  \BibitemOpen
  \bibfield  {author} {\bibinfo {author} {\bibfnamefont {W.}~\bibnamefont
  {Hanke}}\ and\ \bibinfo {author} {\bibfnamefont {L.~J.}\ \bibnamefont
  {Sham}},\ }\href {\doibase 10.1103/physrevb.21.4656} {\bibfield  {journal}
  {\bibinfo  {journal} {Phys. Rev. B}\ }\textbf {\bibinfo {volume} {21}},\
  \bibinfo {pages} {4656} (\bibinfo {year} {1980})}\BibitemShut {NoStop}%
\bibitem [{\citenamefont {Strinati}, \citenamefont {Mattausch},\ and\
  \citenamefont {Hanke}(1982)}]{Strinati1982}%
  \BibitemOpen
  \bibfield  {author} {\bibinfo {author} {\bibfnamefont {G.}~\bibnamefont
  {Strinati}}, \bibinfo {author} {\bibfnamefont {H.~J.}\ \bibnamefont
  {Mattausch}}, \ and\ \bibinfo {author} {\bibfnamefont {W.}~\bibnamefont
  {Hanke}},\ }\href {\doibase 10.1103/physrevb.25.2867} {\bibfield  {journal}
  {\bibinfo  {journal} {Phys. Rev. B}\ }\textbf {\bibinfo {volume} {25}},\
  \bibinfo {pages} {2867} (\bibinfo {year} {1982})}\BibitemShut {NoStop}%
\bibitem [{\citenamefont {Strinati}(1984)}]{Strinati1984}%
  \BibitemOpen
  \bibfield  {author} {\bibinfo {author} {\bibfnamefont {G.}~\bibnamefont
  {Strinati}},\ }\href {\doibase 10.1103/physrevb.29.5718} {\bibfield
  {journal} {\bibinfo  {journal} {Phys. Rev. B}\ }\textbf {\bibinfo {volume}
  {29}},\ \bibinfo {pages} {5718} (\bibinfo {year} {1984})}\BibitemShut
  {NoStop}%
\bibitem [{\citenamefont {Hybertsen}\ and\ \citenamefont
  {Louie}(1985)}]{Hybertsen1985}%
  \BibitemOpen
  \bibfield  {author} {\bibinfo {author} {\bibfnamefont {M.~S.}\ \bibnamefont
  {Hybertsen}}\ and\ \bibinfo {author} {\bibfnamefont {S.~G.}\ \bibnamefont
  {Louie}},\ }\href {\doibase 10.1103/physrevlett.55.1418} {\bibfield
  {journal} {\bibinfo  {journal} {Phys. Rev. Lett.}\ }\textbf {\bibinfo
  {volume} {55}},\ \bibinfo {pages} {1418} (\bibinfo {year}
  {1985})}\BibitemShut {NoStop}%
\bibitem [{\citenamefont {Hybertsen}\ and\ \citenamefont
  {Louie}(1986)}]{Hybertsen1986}%
  \BibitemOpen
  \bibfield  {author} {\bibinfo {author} {\bibfnamefont {M.~S.}\ \bibnamefont
  {Hybertsen}}\ and\ \bibinfo {author} {\bibfnamefont {S.~G.}\ \bibnamefont
  {Louie}},\ }\href {\doibase 10.1103/physrevb.34.5390} {\bibfield  {journal}
  {\bibinfo  {journal} {Phys. Rev. B}\ }\textbf {\bibinfo {volume} {34}},\
  \bibinfo {pages} {5390} (\bibinfo {year} {1986})}\BibitemShut {NoStop}%
\bibitem [{\citenamefont {Albrecht}\ \emph {et~al.}(1998)\citenamefont
  {Albrecht}, \citenamefont {Reining}, \citenamefont {Sole},\ and\
  \citenamefont {Onida}}]{Albrecht1998}%
  \BibitemOpen
  \bibfield  {author} {\bibinfo {author} {\bibfnamefont {S.}~\bibnamefont
  {Albrecht}}, \bibinfo {author} {\bibfnamefont {L.}~\bibnamefont {Reining}},
  \bibinfo {author} {\bibfnamefont {R.~D.}\ \bibnamefont {Sole}}, \ and\
  \bibinfo {author} {\bibfnamefont {G.}~\bibnamefont {Onida}},\ }\href
  {\doibase 10.1002/(sici)1521-396x(199812)170:2<189::aid-pssa189>3.0.co;2-3}
  {\bibfield  {journal} {\bibinfo  {journal} {Phys. Status Solidi}\ }\textbf
  {\bibinfo {volume} {170}},\ \bibinfo {pages} {189} (\bibinfo {year}
  {1998})}\BibitemShut {NoStop}%
\bibitem [{\citenamefont {Rohlfing}\ and\ \citenamefont
  {Louie}(2000)}]{Rohlfing2000}%
  \BibitemOpen
  \bibfield  {author} {\bibinfo {author} {\bibfnamefont {M.}~\bibnamefont
  {Rohlfing}}\ and\ \bibinfo {author} {\bibfnamefont {S.~G.}\ \bibnamefont
  {Louie}},\ }\href {\doibase 10.1103/physrevb.62.4927} {\bibfield  {journal}
  {\bibinfo  {journal} {Phys. Rev. B}\ }\textbf {\bibinfo {volume} {62}},\
  \bibinfo {pages} {4927} (\bibinfo {year} {2000})}\BibitemShut {NoStop}%
\bibitem [{\citenamefont {Tiago}\ and\ \citenamefont
  {Chelikowsky}(2005)}]{Tiago2005}%
  \BibitemOpen
  \bibfield  {author} {\bibinfo {author} {\bibfnamefont {M.~L.}\ \bibnamefont
  {Tiago}}\ and\ \bibinfo {author} {\bibfnamefont {J.~R.}\ \bibnamefont
  {Chelikowsky}},\ }\href {\doibase 10.1016/j.ssc.2005.08.012} {\bibfield
  {journal} {\bibinfo  {journal} {Solid State Commun.}\ }\textbf {\bibinfo
  {volume} {136}},\ \bibinfo {pages} {333} (\bibinfo {year}
  {2005})}\BibitemShut {NoStop}%
\bibitem [{\citenamefont {Faber}\ \emph {et~al.}(2014)\citenamefont {Faber},
  \citenamefont {Boulanger}, \citenamefont {Attaccalite}, \citenamefont
  {Duchemin},\ and\ \citenamefont {Blase}}]{Faber2014}%
  \BibitemOpen
  \bibfield  {author} {\bibinfo {author} {\bibfnamefont {C.}~\bibnamefont
  {Faber}}, \bibinfo {author} {\bibfnamefont {P.}~\bibnamefont {Boulanger}},
  \bibinfo {author} {\bibfnamefont {C.}~\bibnamefont {Attaccalite}}, \bibinfo
  {author} {\bibfnamefont {I.}~\bibnamefont {Duchemin}}, \ and\ \bibinfo
  {author} {\bibfnamefont {X.}~\bibnamefont {Blase}},\ }\href {\doibase
  10.1098/rsta.2013.0271} {\bibfield  {journal} {\bibinfo  {journal} {Philos.
  Trans. Royal Soc. A}\ }\textbf {\bibinfo {volume} {372}},\ \bibinfo {pages}
  {20130271} (\bibinfo {year} {2014})}\BibitemShut {NoStop}%
\bibitem [{\citenamefont {Körbel}\ \emph {et~al.}(2014)\citenamefont
  {Körbel}, \citenamefont {Boulanger}, \citenamefont {Duchemin}, \citenamefont
  {Blase}, \citenamefont {Marques},\ and\ \citenamefont {Botti}}]{Koerbel2014}%
  \BibitemOpen
  \bibfield  {author} {\bibinfo {author} {\bibfnamefont {S.}~\bibnamefont
  {Körbel}}, \bibinfo {author} {\bibfnamefont {P.}~\bibnamefont {Boulanger}},
  \bibinfo {author} {\bibfnamefont {I.}~\bibnamefont {Duchemin}}, \bibinfo
  {author} {\bibfnamefont {X.}~\bibnamefont {Blase}}, \bibinfo {author}
  {\bibfnamefont {M.~A.~L.}\ \bibnamefont {Marques}}, \ and\ \bibinfo {author}
  {\bibfnamefont {S.}~\bibnamefont {Botti}},\ }\href {\doibase
  10.1021/ct5003658} {\bibfield  {journal} {\bibinfo  {journal} {J. Chem.
  Theory Comput.}\ }\textbf {\bibinfo {volume} {10}},\ \bibinfo {pages} {3934}
  (\bibinfo {year} {2014})}\BibitemShut {NoStop}%
\bibitem [{\citenamefont {Bruneval}, \citenamefont {Hamed},\ and\ \citenamefont
  {Neaton}(2015)}]{Bruneval2015}%
  \BibitemOpen
  \bibfield  {author} {\bibinfo {author} {\bibfnamefont {F.}~\bibnamefont
  {Bruneval}}, \bibinfo {author} {\bibfnamefont {S.~M.}\ \bibnamefont {Hamed}},
  \ and\ \bibinfo {author} {\bibfnamefont {J.~B.}\ \bibnamefont {Neaton}},\
  }\href {\doibase 10.1063/1.4922489} {\bibfield  {journal} {\bibinfo
  {journal} {J. Chem. Phys.}\ }\textbf {\bibinfo {volume} {142}},\ \bibinfo
  {pages} {244101} (\bibinfo {year} {2015})}\BibitemShut {NoStop}%
\bibitem [{\citenamefont {Jacquemin}, \citenamefont {Duchemin},\ and\
  \citenamefont {Blase}(2015)}]{Jacquemin2015}%
  \BibitemOpen
  \bibfield  {author} {\bibinfo {author} {\bibfnamefont {D.}~\bibnamefont
  {Jacquemin}}, \bibinfo {author} {\bibfnamefont {I.}~\bibnamefont {Duchemin}},
  \ and\ \bibinfo {author} {\bibfnamefont {X.}~\bibnamefont {Blase}},\ }\href
  {\doibase 10.1021/acs.jctc.5b00304} {\bibfield  {journal} {\bibinfo
  {journal} {J. Chem. Theory Comput.}\ }\textbf {\bibinfo {volume} {11}},\
  \bibinfo {pages} {3290} (\bibinfo {year} {2015})}\BibitemShut {NoStop}%
\bibitem [{\citenamefont {van Setten}\ \emph
  {et~al.}(2015{\natexlab{a}})\citenamefont {van Setten}, \citenamefont
  {Caruso}, \citenamefont {Sharifzadeh}, \citenamefont {Ren}, \citenamefont
  {Scheffler}, \citenamefont {Liu}, \citenamefont {Lischner}, \citenamefont
  {Lin}, \citenamefont {Deslippe}, \citenamefont {Louie}, \citenamefont {Yang},
  \citenamefont {Weigend}, \citenamefont {Neaton}, \citenamefont {Evers},\ and\
  \citenamefont {Rinke}}]{Setten2015}%
  \BibitemOpen
  \bibfield  {author} {\bibinfo {author} {\bibfnamefont {M.~J.}\ \bibnamefont
  {van Setten}}, \bibinfo {author} {\bibfnamefont {F.}~\bibnamefont {Caruso}},
  \bibinfo {author} {\bibfnamefont {S.}~\bibnamefont {Sharifzadeh}}, \bibinfo
  {author} {\bibfnamefont {X.}~\bibnamefont {Ren}}, \bibinfo {author}
  {\bibfnamefont {M.}~\bibnamefont {Scheffler}}, \bibinfo {author}
  {\bibfnamefont {F.}~\bibnamefont {Liu}}, \bibinfo {author} {\bibfnamefont
  {J.}~\bibnamefont {Lischner}}, \bibinfo {author} {\bibfnamefont
  {L.}~\bibnamefont {Lin}}, \bibinfo {author} {\bibfnamefont {J.~R.}\
  \bibnamefont {Deslippe}}, \bibinfo {author} {\bibfnamefont {S.~G.}\
  \bibnamefont {Louie}}, \bibinfo {author} {\bibfnamefont {C.}~\bibnamefont
  {Yang}}, \bibinfo {author} {\bibfnamefont {F.}~\bibnamefont {Weigend}},
  \bibinfo {author} {\bibfnamefont {J.~B.}\ \bibnamefont {Neaton}}, \bibinfo
  {author} {\bibfnamefont {F.}~\bibnamefont {Evers}}, \ and\ \bibinfo {author}
  {\bibfnamefont {P.}~\bibnamefont {Rinke}},\ }\href {\doibase
  10.1021/acs.jctc.5b00453} {\bibfield  {journal} {\bibinfo  {journal} {J.
  Chem. Theory Comput.}\ }\textbf {\bibinfo {volume} {11}},\ \bibinfo {pages}
  {5665} (\bibinfo {year} {2015}{\natexlab{a}})}\BibitemShut {NoStop}%
\bibitem [{\citenamefont {Caruso}\ \emph {et~al.}(2016)\citenamefont {Caruso},
  \citenamefont {Dauth}, \citenamefont {van Setten},\ and\ \citenamefont
  {Rinke}}]{Caruso2016}%
  \BibitemOpen
  \bibfield  {author} {\bibinfo {author} {\bibfnamefont {F.}~\bibnamefont
  {Caruso}}, \bibinfo {author} {\bibfnamefont {M.}~\bibnamefont {Dauth}},
  \bibinfo {author} {\bibfnamefont {M.~J.}\ \bibnamefont {van Setten}}, \ and\
  \bibinfo {author} {\bibfnamefont {P.}~\bibnamefont {Rinke}},\ }\href
  {\doibase 10.1021/acs.jctc.6b00774} {\bibfield  {journal} {\bibinfo
  {journal} {J. Chem. Theory Comput.}\ }\textbf {\bibinfo {volume} {12}},\
  \bibinfo {pages} {5076} (\bibinfo {year} {2016})}\BibitemShut {NoStop}%
\bibitem [{\citenamefont {Knight}\ \emph {et~al.}(2016)\citenamefont {Knight},
  \citenamefont {Wang}, \citenamefont {Gallandi}, \citenamefont
  {Dolgounitcheva}, \citenamefont {Ren}, \citenamefont {Ortiz}, \citenamefont
  {Rinke}, \citenamefont {Körzdörfer},\ and\ \citenamefont
  {Marom}}]{Knight2016}%
  \BibitemOpen
  \bibfield  {author} {\bibinfo {author} {\bibfnamefont {J.~W.}\ \bibnamefont
  {Knight}}, \bibinfo {author} {\bibfnamefont {X.}~\bibnamefont {Wang}},
  \bibinfo {author} {\bibfnamefont {L.}~\bibnamefont {Gallandi}}, \bibinfo
  {author} {\bibfnamefont {O.}~\bibnamefont {Dolgounitcheva}}, \bibinfo
  {author} {\bibfnamefont {X.}~\bibnamefont {Ren}}, \bibinfo {author}
  {\bibfnamefont {J.~V.}\ \bibnamefont {Ortiz}}, \bibinfo {author}
  {\bibfnamefont {P.}~\bibnamefont {Rinke}}, \bibinfo {author} {\bibfnamefont
  {T.}~\bibnamefont {Körzdörfer}}, \ and\ \bibinfo {author} {\bibfnamefont
  {N.}~\bibnamefont {Marom}},\ }\href {\doibase 10.1021/acs.jctc.5b00871}
  {\bibfield  {journal} {\bibinfo  {journal} {J. Chem. Theory Comput.}\
  }\textbf {\bibinfo {volume} {12}},\ \bibinfo {pages} {615} (\bibinfo {year}
  {2016})}\BibitemShut {NoStop}%
\bibitem [{\citenamefont {Rangel}\ \emph {et~al.}(2017)\citenamefont {Rangel},
  \citenamefont {Hamed}, \citenamefont {Bruneval},\ and\ \citenamefont
  {Neaton}}]{Rangel2017}%
  \BibitemOpen
  \bibfield  {author} {\bibinfo {author} {\bibfnamefont {T.}~\bibnamefont
  {Rangel}}, \bibinfo {author} {\bibfnamefont {S.~M.}\ \bibnamefont {Hamed}},
  \bibinfo {author} {\bibfnamefont {F.}~\bibnamefont {Bruneval}}, \ and\
  \bibinfo {author} {\bibfnamefont {J.~B.}\ \bibnamefont {Neaton}},\ }\href
  {\doibase 10.1063/1.4983126} {\bibfield  {journal} {\bibinfo  {journal} {J.
  Chem. Phys.}\ }\textbf {\bibinfo {volume} {146}},\ \bibinfo {pages} {194108}
  (\bibinfo {year} {2017})}\BibitemShut {NoStop}%
\bibitem [{\citenamefont {Blase}, \citenamefont {Duchemin},\ and\ \citenamefont
  {Jacquemin}(2018)}]{Blase2018}%
  \BibitemOpen
  \bibfield  {author} {\bibinfo {author} {\bibfnamefont {X.}~\bibnamefont
  {Blase}}, \bibinfo {author} {\bibfnamefont {I.}~\bibnamefont {Duchemin}}, \
  and\ \bibinfo {author} {\bibfnamefont {D.}~\bibnamefont {Jacquemin}},\ }\href
  {\doibase 10.1039/c7cs00049a} {\bibfield  {journal} {\bibinfo  {journal}
  {Chem. Soc. Rev.}\ }\textbf {\bibinfo {volume} {47}},\ \bibinfo {pages}
  {1022} (\bibinfo {year} {2018})}\BibitemShut {NoStop}%
\bibitem [{\citenamefont {Golze}, \citenamefont {Dvorak},\ and\ \citenamefont
  {Rinke}(2019)}]{Golze2019}%
  \BibitemOpen
  \bibfield  {author} {\bibinfo {author} {\bibfnamefont {D.}~\bibnamefont
  {Golze}}, \bibinfo {author} {\bibfnamefont {M.}~\bibnamefont {Dvorak}}, \
  and\ \bibinfo {author} {\bibfnamefont {P.}~\bibnamefont {Rinke}},\ }\href
  {\doibase 10.3389/fchem.2019.00377} {\bibfield  {journal} {\bibinfo
  {journal} {Front. Chem.}\ }\textbf {\bibinfo {volume} {7}},\ \bibinfo {pages}
  {377} (\bibinfo {year} {2019})}\BibitemShut {NoStop}%
\bibitem [{\citenamefont {Blase}\ \emph {et~al.}(2020)\citenamefont {Blase},
  \citenamefont {Duchemin}, \citenamefont {Jacquemin},\ and\ \citenamefont
  {Loos}}]{Blase2020}%
  \BibitemOpen
  \bibfield  {author} {\bibinfo {author} {\bibfnamefont {X.}~\bibnamefont
  {Blase}}, \bibinfo {author} {\bibfnamefont {I.}~\bibnamefont {Duchemin}},
  \bibinfo {author} {\bibfnamefont {D.}~\bibnamefont {Jacquemin}}, \ and\
  \bibinfo {author} {\bibfnamefont {P.-F.}\ \bibnamefont {Loos}},\ }\href
  {\doibase 10.1021/acs.jpclett.0c01875} {\bibfield  {journal} {\bibinfo
  {journal} {J. Phys. Chem. Lett.}\ }\textbf {\bibinfo {volume} {11}},\
  \bibinfo {pages} {7371} (\bibinfo {year} {2020})}\BibitemShut {NoStop}%
\bibitem [{\citenamefont {Foerster}, \citenamefont {Koval},\ and\ \citenamefont
  {S{\'{a}}nchez-Portal}(2011)}]{Foerster2011}%
  \BibitemOpen
  \bibfield  {author} {\bibinfo {author} {\bibfnamefont {D.}~\bibnamefont
  {Foerster}}, \bibinfo {author} {\bibfnamefont {P.}~\bibnamefont {Koval}}, \
  and\ \bibinfo {author} {\bibfnamefont {D.}~\bibnamefont
  {S{\'{a}}nchez-Portal}},\ }\href {\doibase 10.1063/1.3624731} {\bibfield
  {journal} {\bibinfo  {journal} {J. Chem. Phys.}\ }\textbf {\bibinfo {volume}
  {135}},\ \bibinfo {pages} {074105} (\bibinfo {year} {2011})}\BibitemShut
  {NoStop}%
\bibitem [{\citenamefont {Deslippe}\ \emph {et~al.}(2012)\citenamefont
  {Deslippe}, \citenamefont {Samsonidze}, \citenamefont {Strubbe},
  \citenamefont {Jain}, \citenamefont {Cohen},\ and\ \citenamefont
  {Louie}}]{Deslippe2012}%
  \BibitemOpen
  \bibfield  {author} {\bibinfo {author} {\bibfnamefont {J.}~\bibnamefont
  {Deslippe}}, \bibinfo {author} {\bibfnamefont {G.}~\bibnamefont
  {Samsonidze}}, \bibinfo {author} {\bibfnamefont {D.~A.}\ \bibnamefont
  {Strubbe}}, \bibinfo {author} {\bibfnamefont {M.}~\bibnamefont {Jain}},
  \bibinfo {author} {\bibfnamefont {M.~L.}\ \bibnamefont {Cohen}}, \ and\
  \bibinfo {author} {\bibfnamefont {S.~G.}\ \bibnamefont {Louie}},\ }\href
  {\doibase 10.1016/j.cpc.2011.12.006} {\bibfield  {journal} {\bibinfo
  {journal} {Comput. Phys. Commun.}\ }\textbf {\bibinfo {volume} {183}},\
  \bibinfo {pages} {1269} (\bibinfo {year} {2012})}\BibitemShut {NoStop}%
\bibitem [{\citenamefont {van Setten}, \citenamefont {Weigend},\ and\
  \citenamefont {Evers}(2013)}]{setten2013}%
  \BibitemOpen
  \bibfield  {author} {\bibinfo {author} {\bibfnamefont {M.~J.}\ \bibnamefont
  {van Setten}}, \bibinfo {author} {\bibfnamefont {F.}~\bibnamefont {Weigend}},
  \ and\ \bibinfo {author} {\bibfnamefont {F.}~\bibnamefont {Evers}},\ }\href
  {\doibase 10.1021/ct300648t} {\bibfield  {journal} {\bibinfo  {journal} {J.
  Chem. Theory Comput.}\ }\textbf {\bibinfo {volume} {9}},\ \bibinfo {pages}
  {232} (\bibinfo {year} {2013})}\BibitemShut {NoStop}%
\bibitem [{\citenamefont {Govoni}\ and\ \citenamefont
  {Galli}(2015)}]{Govoni2015}%
  \BibitemOpen
  \bibfield  {author} {\bibinfo {author} {\bibfnamefont {M.}~\bibnamefont
  {Govoni}}\ and\ \bibinfo {author} {\bibfnamefont {G.}~\bibnamefont {Galli}},\
  }\href {\doibase 10.1021/ct500958p} {\bibfield  {journal} {\bibinfo
  {journal} {J. Chem. Theory Comput.}\ }\textbf {\bibinfo {volume} {11}},\
  \bibinfo {pages} {2680} (\bibinfo {year} {2015})}\BibitemShut {NoStop}%
\bibitem [{\citenamefont {Ljungberg}\ \emph {et~al.}(2015)\citenamefont
  {Ljungberg}, \citenamefont {Koval}, \citenamefont {Ferrari}, \citenamefont
  {Foerster},\ and\ \citenamefont {S{\'{a}}nchez-Portal}}]{Ljungberg2015}%
  \BibitemOpen
  \bibfield  {author} {\bibinfo {author} {\bibfnamefont {M.~P.}\ \bibnamefont
  {Ljungberg}}, \bibinfo {author} {\bibfnamefont {P.}~\bibnamefont {Koval}},
  \bibinfo {author} {\bibfnamefont {F.}~\bibnamefont {Ferrari}}, \bibinfo
  {author} {\bibfnamefont {D.}~\bibnamefont {Foerster}}, \ and\ \bibinfo
  {author} {\bibfnamefont {D.}~\bibnamefont {S{\'{a}}nchez-Portal}},\ }\href
  {\doibase 10.1103/physrevb.92.075422} {\bibfield  {journal} {\bibinfo
  {journal} {Phys. Rev. B}\ }\textbf {\bibinfo {volume} {92}},\ \bibinfo
  {pages} {075422} (\bibinfo {year} {2015})}\BibitemShut {NoStop}%
\bibitem [{\citenamefont {Krause}\ and\ \citenamefont
  {Klopper}(2016)}]{Krause2016}%
  \BibitemOpen
  \bibfield  {author} {\bibinfo {author} {\bibfnamefont {K.}~\bibnamefont
  {Krause}}\ and\ \bibinfo {author} {\bibfnamefont {W.}~\bibnamefont
  {Klopper}},\ }\href {\doibase 10.1002/jcc.24688} {\bibfield  {journal}
  {\bibinfo  {journal} {J. Comput. Chem.}\ }\textbf {\bibinfo {volume} {38}},\
  \bibinfo {pages} {383} (\bibinfo {year} {2016})}\BibitemShut {NoStop}%
\bibitem [{\citenamefont {Bruneval}\ \emph {et~al.}(2016)\citenamefont
  {Bruneval}, \citenamefont {Rangel}, \citenamefont {Hamed}, \citenamefont
  {Shao}, \citenamefont {Yang},\ and\ \citenamefont {Neaton}}]{Bruneval2016}%
  \BibitemOpen
  \bibfield  {author} {\bibinfo {author} {\bibfnamefont {F.}~\bibnamefont
  {Bruneval}}, \bibinfo {author} {\bibfnamefont {T.}~\bibnamefont {Rangel}},
  \bibinfo {author} {\bibfnamefont {S.~M.}\ \bibnamefont {Hamed}}, \bibinfo
  {author} {\bibfnamefont {M.}~\bibnamefont {Shao}}, \bibinfo {author}
  {\bibfnamefont {C.}~\bibnamefont {Yang}}, \ and\ \bibinfo {author}
  {\bibfnamefont {J.~B.}\ \bibnamefont {Neaton}},\ }\href {\doibase
  10.1016/j.cpc.2016.06.019} {\bibfield  {journal} {\bibinfo  {journal}
  {Comput. Phys. Commun.}\ }\textbf {\bibinfo {volume} {208}},\ \bibinfo
  {pages} {149} (\bibinfo {year} {2016})}\BibitemShut {NoStop}%
\bibitem [{\citenamefont {Wilhelm}\ \emph {et~al.}(2018)\citenamefont
  {Wilhelm}, \citenamefont {Golze}, \citenamefont {Talirz}, \citenamefont
  {Hutter},\ and\ \citenamefont {Pignedoli}}]{Wilhelm2018}%
  \BibitemOpen
  \bibfield  {author} {\bibinfo {author} {\bibfnamefont {J.}~\bibnamefont
  {Wilhelm}}, \bibinfo {author} {\bibfnamefont {D.}~\bibnamefont {Golze}},
  \bibinfo {author} {\bibfnamefont {L.}~\bibnamefont {Talirz}}, \bibinfo
  {author} {\bibfnamefont {J.}~\bibnamefont {Hutter}}, \ and\ \bibinfo {author}
  {\bibfnamefont {C.~A.}\ \bibnamefont {Pignedoli}},\ }\href {\doibase
  10.1021/acs.jpclett.7b02740} {\bibfield  {journal} {\bibinfo  {journal} {J.
  Phys. Chem. Lett.}\ }\textbf {\bibinfo {volume} {9}},\ \bibinfo {pages} {306}
  (\bibinfo {year} {2018})}\BibitemShut {NoStop}%
\bibitem [{\citenamefont {Koval}\ \emph {et~al.}(2019)\citenamefont {Koval},
  \citenamefont {Ljungberg}, \citenamefont {Müller},\ and\ \citenamefont
  {S{\'{a}}nchez-Portal}}]{Koval2019}%
  \BibitemOpen
  \bibfield  {author} {\bibinfo {author} {\bibfnamefont {P.}~\bibnamefont
  {Koval}}, \bibinfo {author} {\bibfnamefont {M.~P.}\ \bibnamefont
  {Ljungberg}}, \bibinfo {author} {\bibfnamefont {M.}~\bibnamefont {Müller}},
  \ and\ \bibinfo {author} {\bibfnamefont {D.}~\bibnamefont
  {S{\'{a}}nchez-Portal}},\ }\href {\doibase 10.1021/acs.jctc.9b00436}
  {\bibfield  {journal} {\bibinfo  {journal} {J. Chem. Theory Comput.}\
  }\textbf {\bibinfo {volume} {15}},\ \bibinfo {pages} {4564} (\bibinfo {year}
  {2019})}\BibitemShut {NoStop}%
\bibitem [{\citenamefont {Liu}\ \emph {et~al.}(2020)\citenamefont {Liu},
  \citenamefont {Kloppenburg}, \citenamefont {Yao}, \citenamefont {Ren},
  \citenamefont {Appel}, \citenamefont {Kanai},\ and\ \citenamefont
  {Blum}}]{Liu2020}%
  \BibitemOpen
  \bibfield  {author} {\bibinfo {author} {\bibfnamefont {C.}~\bibnamefont
  {Liu}}, \bibinfo {author} {\bibfnamefont {J.}~\bibnamefont {Kloppenburg}},
  \bibinfo {author} {\bibfnamefont {Y.}~\bibnamefont {Yao}}, \bibinfo {author}
  {\bibfnamefont {X.}~\bibnamefont {Ren}}, \bibinfo {author} {\bibfnamefont
  {H.}~\bibnamefont {Appel}}, \bibinfo {author} {\bibfnamefont
  {Y.}~\bibnamefont {Kanai}}, \ and\ \bibinfo {author} {\bibfnamefont
  {V.}~\bibnamefont {Blum}},\ }\href {\doibase 10.1063/1.5123290} {\bibfield
  {journal} {\bibinfo  {journal} {J. Chem. Phys.}\ }\textbf {\bibinfo {volume}
  {152}},\ \bibinfo {pages} {044105} (\bibinfo {year} {2020})}\BibitemShut
  {NoStop}%
\bibitem [{\citenamefont {Zhu}\ and\ \citenamefont
  {Chan}(2021{\natexlab{a}})}]{Zhu2021}%
  \BibitemOpen
  \bibfield  {author} {\bibinfo {author} {\bibfnamefont {T.}~\bibnamefont
  {Zhu}}\ and\ \bibinfo {author} {\bibfnamefont {G.~K.-L.}\ \bibnamefont
  {Chan}},\ }\href {\doibase 10.1021/acs.jctc.0c00704} {\bibfield  {journal}
  {\bibinfo  {journal} {J. Chem. Theory Comput.}\ }\textbf {\bibinfo {volume}
  {17}},\ \bibinfo {pages} {727} (\bibinfo {year}
  {2021}{\natexlab{a}})}\BibitemShut {NoStop}%
\bibitem [{\citenamefont {Bintrim}\ and\ \citenamefont
  {Berkelbach}(2021)}]{Bintrim2021}%
  \BibitemOpen
  \bibfield  {author} {\bibinfo {author} {\bibfnamefont {S.~J.}\ \bibnamefont
  {Bintrim}}\ and\ \bibinfo {author} {\bibfnamefont {T.~C.}\ \bibnamefont
  {Berkelbach}},\ }\href {\doibase 10.1063/5.0035141} {\bibfield  {journal}
  {\bibinfo  {journal} {J. Chem. Phys.}\ }\textbf {\bibinfo {volume} {154}},\
  \bibinfo {pages} {041101} (\bibinfo {year} {2021})}\BibitemShut {NoStop}%
\bibitem [{\citenamefont {Elstner}\ \emph {et~al.}(1998)\citenamefont
  {Elstner}, \citenamefont {Porezag}, \citenamefont {Jungnickel}, \citenamefont
  {Elsner}, \citenamefont {Haugk}, \citenamefont {Frauenheim}, \citenamefont
  {Suhai},\ and\ \citenamefont {Seifert}}]{Elstner1998}%
  \BibitemOpen
  \bibfield  {author} {\bibinfo {author} {\bibfnamefont {M.}~\bibnamefont
  {Elstner}}, \bibinfo {author} {\bibfnamefont {D.}~\bibnamefont {Porezag}},
  \bibinfo {author} {\bibfnamefont {G.}~\bibnamefont {Jungnickel}}, \bibinfo
  {author} {\bibfnamefont {J.}~\bibnamefont {Elsner}}, \bibinfo {author}
  {\bibfnamefont {M.}~\bibnamefont {Haugk}}, \bibinfo {author} {\bibfnamefont
  {T.}~\bibnamefont {Frauenheim}}, \bibinfo {author} {\bibfnamefont
  {S.}~\bibnamefont {Suhai}}, \ and\ \bibinfo {author} {\bibfnamefont
  {G.}~\bibnamefont {Seifert}},\ }\href {\doibase 10.1103/physrevb.58.7260}
  {\bibfield  {journal} {\bibinfo  {journal} {Phys. Rev. B}\ }\textbf {\bibinfo
  {volume} {58}},\ \bibinfo {pages} {7260} (\bibinfo {year}
  {1998})}\BibitemShut {NoStop}%
\bibitem [{\citenamefont {Hourahine}\ \emph {et~al.}(2020)\citenamefont
  {Hourahine}, \citenamefont {Aradi}, \citenamefont {Blum}, \citenamefont
  {Bonaf{\'e}}, \citenamefont {Buccheri}, \citenamefont {Camacho},
  \citenamefont {Cevallos}, \citenamefont {Deshaye}, \citenamefont
  {Dumitric{\u{a}}}, \citenamefont {Dominguez} \emph
  {et~al.}}]{hourahine2020dftb+}%
  \BibitemOpen
  \bibfield  {author} {\bibinfo {author} {\bibfnamefont {B.}~\bibnamefont
  {Hourahine}}, \bibinfo {author} {\bibfnamefont {B.}~\bibnamefont {Aradi}},
  \bibinfo {author} {\bibfnamefont {V.}~\bibnamefont {Blum}}, \bibinfo {author}
  {\bibfnamefont {F.}~\bibnamefont {Bonaf{\'e}}}, \bibinfo {author}
  {\bibfnamefont {A.}~\bibnamefont {Buccheri}}, \bibinfo {author}
  {\bibfnamefont {C.}~\bibnamefont {Camacho}}, \bibinfo {author} {\bibfnamefont
  {C.}~\bibnamefont {Cevallos}}, \bibinfo {author} {\bibfnamefont
  {M.}~\bibnamefont {Deshaye}}, \bibinfo {author} {\bibfnamefont
  {T.}~\bibnamefont {Dumitric{\u{a}}}}, \bibinfo {author} {\bibfnamefont
  {A.}~\bibnamefont {Dominguez}},  \emph {et~al.},\ }\href {\doibase
  10.1063/1.5143190} {\bibfield  {journal} {\bibinfo  {journal} {J. Chem.
  Phys.}\ }\textbf {\bibinfo {volume} {152}},\ \bibinfo {pages} {124101}
  (\bibinfo {year} {2020})}\BibitemShut {NoStop}%
\bibitem [{\citenamefont {Grimme}, \citenamefont {Bannwarth},\ and\
  \citenamefont {Shushkov}(2017)}]{Grimme2017}%
  \BibitemOpen
  \bibfield  {author} {\bibinfo {author} {\bibfnamefont {S.}~\bibnamefont
  {Grimme}}, \bibinfo {author} {\bibfnamefont {C.}~\bibnamefont {Bannwarth}}, \
  and\ \bibinfo {author} {\bibfnamefont {P.}~\bibnamefont {Shushkov}},\ }\href
  {\doibase 10.1021/acs.jctc.7b00118} {\bibfield  {journal} {\bibinfo
  {journal} {J. Chem. Theory Comput.}\ }\textbf {\bibinfo {volume} {13}},\
  \bibinfo {pages} {1989} (\bibinfo {year} {2017})}\BibitemShut {NoStop}%
\bibitem [{\citenamefont {Bannwarth}, \citenamefont {Ehlert},\ and\
  \citenamefont {Grimme}(2019)}]{bannwarth2019gfn2}%
  \BibitemOpen
  \bibfield  {author} {\bibinfo {author} {\bibfnamefont {C.}~\bibnamefont
  {Bannwarth}}, \bibinfo {author} {\bibfnamefont {S.}~\bibnamefont {Ehlert}}, \
  and\ \bibinfo {author} {\bibfnamefont {S.}~\bibnamefont {Grimme}},\ }\href
  {\doibase 10.1021/acs.jctc.8b01176} {\bibfield  {journal} {\bibinfo
  {journal} {J. Chem. Theory Comput.}\ }\textbf {\bibinfo {volume} {15}},\
  \bibinfo {pages} {1652} (\bibinfo {year} {2019})}\BibitemShut {NoStop}%
\bibitem [{\citenamefont {Bannwarth}\ \emph {et~al.}(2020)\citenamefont
  {Bannwarth}, \citenamefont {Caldeweyher}, \citenamefont {Ehlert},
  \citenamefont {Hansen}, \citenamefont {Pracht}, \citenamefont {Seibert},
  \citenamefont {Spicher},\ and\ \citenamefont {Grimme}}]{Bannwarth2020}%
  \BibitemOpen
  \bibfield  {author} {\bibinfo {author} {\bibfnamefont {C.}~\bibnamefont
  {Bannwarth}}, \bibinfo {author} {\bibfnamefont {E.}~\bibnamefont
  {Caldeweyher}}, \bibinfo {author} {\bibfnamefont {S.}~\bibnamefont {Ehlert}},
  \bibinfo {author} {\bibfnamefont {A.}~\bibnamefont {Hansen}}, \bibinfo
  {author} {\bibfnamefont {P.}~\bibnamefont {Pracht}}, \bibinfo {author}
  {\bibfnamefont {J.}~\bibnamefont {Seibert}}, \bibinfo {author} {\bibfnamefont
  {S.}~\bibnamefont {Spicher}}, \ and\ \bibinfo {author} {\bibfnamefont
  {S.}~\bibnamefont {Grimme}},\ }\href {\doibase 10.1002/wcms.1493} {\bibfield
  {journal} {\bibinfo  {journal} {{WIREs} Comput. Mol. Sci.}\ }\textbf
  {\bibinfo {volume} {11}} (\bibinfo {year} {2020}),\
  10.1002/wcms.1493}\BibitemShut {NoStop}%
\bibitem [{\citenamefont {Niehaus}\ \emph {et~al.}(2001)\citenamefont
  {Niehaus}, \citenamefont {Suhai}, \citenamefont {Della~Sala}, \citenamefont
  {Lugli}, \citenamefont {Elstner}, \citenamefont {Seifert},\ and\
  \citenamefont {Frauenheim}}]{niehaus2001tight}%
  \BibitemOpen
  \bibfield  {author} {\bibinfo {author} {\bibfnamefont {T.~A.}\ \bibnamefont
  {Niehaus}}, \bibinfo {author} {\bibfnamefont {S.}~\bibnamefont {Suhai}},
  \bibinfo {author} {\bibfnamefont {F.}~\bibnamefont {Della~Sala}}, \bibinfo
  {author} {\bibfnamefont {P.}~\bibnamefont {Lugli}}, \bibinfo {author}
  {\bibfnamefont {M.}~\bibnamefont {Elstner}}, \bibinfo {author} {\bibfnamefont
  {G.}~\bibnamefont {Seifert}}, \ and\ \bibinfo {author} {\bibfnamefont
  {T.}~\bibnamefont {Frauenheim}},\ }\href {\doibase
  10.1103/PhysRevB.63.085108} {\bibfield  {journal} {\bibinfo  {journal} {Phys.
  Rev. B}\ }\textbf {\bibinfo {volume} {63}},\ \bibinfo {pages} {085108}
  (\bibinfo {year} {2001})}\BibitemShut {NoStop}%
\bibitem [{\citenamefont {Trani}\ \emph {et~al.}(2011)\citenamefont {Trani},
  \citenamefont {Scalmani}, \citenamefont {Zheng}, \citenamefont {Carnimeo},
  \citenamefont {Frisch},\ and\ \citenamefont {Barone}}]{Trani2011}%
  \BibitemOpen
  \bibfield  {author} {\bibinfo {author} {\bibfnamefont {F.}~\bibnamefont
  {Trani}}, \bibinfo {author} {\bibfnamefont {G.}~\bibnamefont {Scalmani}},
  \bibinfo {author} {\bibfnamefont {G.}~\bibnamefont {Zheng}}, \bibinfo
  {author} {\bibfnamefont {I.}~\bibnamefont {Carnimeo}}, \bibinfo {author}
  {\bibfnamefont {M.~J.}\ \bibnamefont {Frisch}}, \ and\ \bibinfo {author}
  {\bibfnamefont {V.}~\bibnamefont {Barone}},\ }\href {\doibase
  10.1021/ct200461y} {\bibfield  {journal} {\bibinfo  {journal} {J. Chem.
  Theory Comput.}\ }\textbf {\bibinfo {volume} {7}},\ \bibinfo {pages} {3304}
  (\bibinfo {year} {2011})}\BibitemShut {NoStop}%
\bibitem [{\citenamefont {Grimme}\ and\ \citenamefont
  {Bannwarth}(2016)}]{grimme2016ultra}%
  \BibitemOpen
  \bibfield  {author} {\bibinfo {author} {\bibfnamefont {S.}~\bibnamefont
  {Grimme}}\ and\ \bibinfo {author} {\bibfnamefont {C.}~\bibnamefont
  {Bannwarth}},\ }\href {\doibase 10.1063/1.4959605} {\bibfield  {journal}
  {\bibinfo  {journal} {J. Chem. Phys.}\ }\textbf {\bibinfo {volume} {145}},\
  \bibinfo {pages} {054103} (\bibinfo {year} {2016})}\BibitemShut {NoStop}%
\bibitem [{\citenamefont {Grimme}(2013)}]{grimme2013simplified}%
  \BibitemOpen
  \bibfield  {author} {\bibinfo {author} {\bibfnamefont {S.}~\bibnamefont
  {Grimme}},\ }\href {\doibase 10.1063/1.4811331} {\bibfield  {journal}
  {\bibinfo  {journal} {J. Chem. Phys.}\ }\textbf {\bibinfo {volume} {138}},\
  \bibinfo {pages} {244104} (\bibinfo {year} {2013})}\BibitemShut {NoStop}%
\bibitem [{\citenamefont {Nishimoto}\ and\ \citenamefont
  {Mataga}(1957)}]{nishimoto1957electronic}%
  \BibitemOpen
  \bibfield  {author} {\bibinfo {author} {\bibfnamefont {K.}~\bibnamefont
  {Nishimoto}}\ and\ \bibinfo {author} {\bibfnamefont {N.}~\bibnamefont
  {Mataga}},\ }\href {\doibase 10.1524/zpch.1957.12.5_6.335} {\bibfield
  {journal} {\bibinfo  {journal} {Z. Phys. Chem.}\ }\textbf {\bibinfo {volume}
  {12}},\ \bibinfo {pages} {335} (\bibinfo {year} {1957})}\BibitemShut
  {NoStop}%
\bibitem [{\citenamefont {Ohno}(1964)}]{ohno1964some}%
  \BibitemOpen
  \bibfield  {author} {\bibinfo {author} {\bibfnamefont {K.}~\bibnamefont
  {Ohno}},\ }\href {\doibase 10.1007/BF00528281} {\bibfield  {journal}
  {\bibinfo  {journal} {Theor. Chim. Acta}\ }\textbf {\bibinfo {volume} {2}},\
  \bibinfo {pages} {219} (\bibinfo {year} {1964})}\BibitemShut {NoStop}%
\bibitem [{\citenamefont {Klopman}(1964)}]{klopman1964semiempirical}%
  \BibitemOpen
  \bibfield  {author} {\bibinfo {author} {\bibfnamefont {G.}~\bibnamefont
  {Klopman}},\ }\href {\doibase 10.1021/ja01075a008} {\bibfield  {journal}
  {\bibinfo  {journal} {J. Am. Chem. Soc.}\ }\textbf {\bibinfo {volume} {86}},\
  \bibinfo {pages} {4550} (\bibinfo {year} {1964})}\BibitemShut {NoStop}%
\bibitem [{\citenamefont {Whitten}(1973)}]{Whitten1973}%
  \BibitemOpen
  \bibfield  {author} {\bibinfo {author} {\bibfnamefont {J.~L.}\ \bibnamefont
  {Whitten}},\ }\href {\doibase 10.1063/1.1679012} {\bibfield  {journal}
  {\bibinfo  {journal} {J. Chem. Phys.}\ }\textbf {\bibinfo {volume} {58}},\
  \bibinfo {pages} {4496} (\bibinfo {year} {1973})}\BibitemShut {NoStop}%
\bibitem [{\citenamefont {Vahtras}, \citenamefont {Almlöf},\ and\
  \citenamefont {Feyereisen}(1993)}]{Vahtras1993}%
  \BibitemOpen
  \bibfield  {author} {\bibinfo {author} {\bibfnamefont {O.}~\bibnamefont
  {Vahtras}}, \bibinfo {author} {\bibfnamefont {J.}~\bibnamefont {Almlöf}}, \
  and\ \bibinfo {author} {\bibfnamefont {M.}~\bibnamefont {Feyereisen}},\
  }\href {\doibase 10.1016/0009-2614(93)89151-7} {\bibfield  {journal}
  {\bibinfo  {journal} {Chem. Phys. Lett.}\ }\textbf {\bibinfo {volume}
  {213}},\ \bibinfo {pages} {514} (\bibinfo {year} {1993})}\BibitemShut
  {NoStop}%
\bibitem [{\citenamefont {Feyereisen}, \citenamefont {Fitzgerald},\ and\
  \citenamefont {Komornicki}(1993)}]{Feyereisen1993}%
  \BibitemOpen
  \bibfield  {author} {\bibinfo {author} {\bibfnamefont {M.}~\bibnamefont
  {Feyereisen}}, \bibinfo {author} {\bibfnamefont {G.}~\bibnamefont
  {Fitzgerald}}, \ and\ \bibinfo {author} {\bibfnamefont {A.}~\bibnamefont
  {Komornicki}},\ }\href {\doibase 10.1016/0009-2614(93)87156-w} {\bibfield
  {journal} {\bibinfo  {journal} {Chem. Phys. Lett.}\ }\textbf {\bibinfo
  {volume} {208}},\ \bibinfo {pages} {359} (\bibinfo {year}
  {1993})}\BibitemShut {NoStop}%
\bibitem [{\citenamefont {Werner}, \citenamefont {Manby},\ and\ \citenamefont
  {Knowles}(2003)}]{Werner2003}%
  \BibitemOpen
  \bibfield  {author} {\bibinfo {author} {\bibfnamefont {H.-J.}\ \bibnamefont
  {Werner}}, \bibinfo {author} {\bibfnamefont {F.~R.}\ \bibnamefont {Manby}}, \
  and\ \bibinfo {author} {\bibfnamefont {P.~J.}\ \bibnamefont {Knowles}},\
  }\href {\doibase 10.1063/1.1564816} {\bibfield  {journal} {\bibinfo
  {journal} {J. Chem. Phys.}\ }\textbf {\bibinfo {volume} {118}},\ \bibinfo
  {pages} {8149} (\bibinfo {year} {2003})}\BibitemShut {NoStop}%
\bibitem [{\citenamefont {Ren}\ \emph {et~al.}(2012)\citenamefont {Ren},
  \citenamefont {Rinke}, \citenamefont {Blum}, \citenamefont {Wieferink},
  \citenamefont {Tkatchenko}, \citenamefont {Sanfilippo}, \citenamefont
  {Reuter},\ and\ \citenamefont {Scheffler}}]{Ren2012}%
  \BibitemOpen
  \bibfield  {author} {\bibinfo {author} {\bibfnamefont {X.}~\bibnamefont
  {Ren}}, \bibinfo {author} {\bibfnamefont {P.}~\bibnamefont {Rinke}}, \bibinfo
  {author} {\bibfnamefont {V.}~\bibnamefont {Blum}}, \bibinfo {author}
  {\bibfnamefont {J.}~\bibnamefont {Wieferink}}, \bibinfo {author}
  {\bibfnamefont {A.}~\bibnamefont {Tkatchenko}}, \bibinfo {author}
  {\bibfnamefont {A.}~\bibnamefont {Sanfilippo}}, \bibinfo {author}
  {\bibfnamefont {K.}~\bibnamefont {Reuter}}, \ and\ \bibinfo {author}
  {\bibfnamefont {M.}~\bibnamefont {Scheffler}},\ }\href {\doibase
  10.1088/1367-2630/14/5/053020} {\bibfield  {journal} {\bibinfo  {journal}
  {New J. Phys.}\ }\textbf {\bibinfo {volume} {14}},\ \bibinfo {pages} {053020}
  (\bibinfo {year} {2012})}\BibitemShut {NoStop}%
\bibitem [{\citenamefont {Wilhelm}, \citenamefont {Ben},\ and\ \citenamefont
  {Hutter}(2016)}]{Wilhelm2016}%
  \BibitemOpen
  \bibfield  {author} {\bibinfo {author} {\bibfnamefont {J.}~\bibnamefont
  {Wilhelm}}, \bibinfo {author} {\bibfnamefont {M.~D.}\ \bibnamefont {Ben}}, \
  and\ \bibinfo {author} {\bibfnamefont {J.}~\bibnamefont {Hutter}},\ }\href
  {\doibase 10.1021/acs.jctc.6b00380} {\bibfield  {journal} {\bibinfo
  {journal} {J. Chem. Theory Comput.}\ }\textbf {\bibinfo {volume} {12}},\
  \bibinfo {pages} {3623} (\bibinfo {year} {2016})}\BibitemShut {NoStop}%
\bibitem [{\citenamefont {Hohenstein}, \citenamefont {Parrish},\ and\
  \citenamefont {Mart{\'{\i}}nez}(2012)}]{Hohenstein2012}%
  \BibitemOpen
  \bibfield  {author} {\bibinfo {author} {\bibfnamefont {E.~G.}\ \bibnamefont
  {Hohenstein}}, \bibinfo {author} {\bibfnamefont {R.~M.}\ \bibnamefont
  {Parrish}}, \ and\ \bibinfo {author} {\bibfnamefont {T.~J.}\ \bibnamefont
  {Mart{\'{\i}}nez}},\ }\href {\doibase 10.1063/1.4732310} {\bibfield
  {journal} {\bibinfo  {journal} {J. Chem. Phys.}\ }\textbf {\bibinfo {volume}
  {137}},\ \bibinfo {pages} {044103} (\bibinfo {year} {2012})}\BibitemShut
  {NoStop}%
\bibitem [{\citenamefont {von~der Linden}\ and\ \citenamefont
  {Horsch}(1988)}]{von1988precise}%
  \BibitemOpen
  \bibfield  {author} {\bibinfo {author} {\bibfnamefont {W.}~\bibnamefont
  {von~der Linden}}\ and\ \bibinfo {author} {\bibfnamefont {P.}~\bibnamefont
  {Horsch}},\ }\href {\doibase 10.1103/PhysRevB.37.8351} {\bibfield  {journal}
  {\bibinfo  {journal} {Phys. Rev. B}\ }\textbf {\bibinfo {volume} {37}},\
  \bibinfo {pages} {8351} (\bibinfo {year} {1988})}\BibitemShut {NoStop}%
\bibitem [{\citenamefont {Larson}, \citenamefont {Dvorak},\ and\ \citenamefont
  {Wu}(2013)}]{Larson2013}%
  \BibitemOpen
  \bibfield  {author} {\bibinfo {author} {\bibfnamefont {P.}~\bibnamefont
  {Larson}}, \bibinfo {author} {\bibfnamefont {M.}~\bibnamefont {Dvorak}}, \
  and\ \bibinfo {author} {\bibfnamefont {Z.}~\bibnamefont {Wu}},\ }\href
  {\doibase 10.1103/physrevb.88.125205} {\bibfield  {journal} {\bibinfo
  {journal} {Phys. Rev. B}\ }\textbf {\bibinfo {volume} {88}},\ \bibinfo
  {pages} {125205} (\bibinfo {year} {2013})}\BibitemShut {NoStop}%
\bibitem [{\citenamefont {Jacquemin}\ \emph {et~al.}(2017)\citenamefont
  {Jacquemin}, \citenamefont {Duchemin}, \citenamefont {Blondel},\ and\
  \citenamefont {Blase}}]{Jacquemin2017}%
  \BibitemOpen
  \bibfield  {author} {\bibinfo {author} {\bibfnamefont {D.}~\bibnamefont
  {Jacquemin}}, \bibinfo {author} {\bibfnamefont {I.}~\bibnamefont {Duchemin}},
  \bibinfo {author} {\bibfnamefont {A.}~\bibnamefont {Blondel}}, \ and\
  \bibinfo {author} {\bibfnamefont {X.}~\bibnamefont {Blase}},\ }\href
  {\doibase 10.1021/acs.jctc.6b01169} {\bibfield  {journal} {\bibinfo
  {journal} {J. Chem. Theory Comput.}\ }\textbf {\bibinfo {volume} {13}},\
  \bibinfo {pages} {767} (\bibinfo {year} {2017})}\BibitemShut {NoStop}%
\bibitem [{\citenamefont {Sun}\ \emph {et~al.}(2018)\citenamefont {Sun},
  \citenamefont {Berkelbach}, \citenamefont {Blunt}, \citenamefont {Booth},
  \citenamefont {Guo}, \citenamefont {Li}, \citenamefont {Liu}, \citenamefont
  {McClain}, \citenamefont {Sayfutyarova}, \citenamefont {Sharma} \emph
  {et~al.}}]{sun2018pyscf}%
  \BibitemOpen
  \bibfield  {author} {\bibinfo {author} {\bibfnamefont {Q.}~\bibnamefont
  {Sun}}, \bibinfo {author} {\bibfnamefont {T.~C.}\ \bibnamefont {Berkelbach}},
  \bibinfo {author} {\bibfnamefont {N.~S.}\ \bibnamefont {Blunt}}, \bibinfo
  {author} {\bibfnamefont {G.~H.}\ \bibnamefont {Booth}}, \bibinfo {author}
  {\bibfnamefont {S.}~\bibnamefont {Guo}}, \bibinfo {author} {\bibfnamefont
  {Z.}~\bibnamefont {Li}}, \bibinfo {author} {\bibfnamefont {J.}~\bibnamefont
  {Liu}}, \bibinfo {author} {\bibfnamefont {J.~D.}\ \bibnamefont {McClain}},
  \bibinfo {author} {\bibfnamefont {E.~R.}\ \bibnamefont {Sayfutyarova}},
  \bibinfo {author} {\bibfnamefont {S.}~\bibnamefont {Sharma}},  \emph
  {et~al.},\ }\href {\doibase 10.1002/wcms.1340} {\bibfield  {journal}
  {\bibinfo  {journal} {WIREs Comput. Mol. Sci.}\ }\textbf {\bibinfo {volume}
  {8}},\ \bibinfo {pages} {e1340} (\bibinfo {year} {2018})}\BibitemShut
  {NoStop}%
\bibitem [{\citenamefont {Sun}\ \emph {et~al.}(2020)\citenamefont {Sun},
  \citenamefont {Zhang}, \citenamefont {Banerjee}, \citenamefont {Bao},
  \citenamefont {Barbry}, \citenamefont {Blunt}, \citenamefont {Bogdanov},
  \citenamefont {Booth}, \citenamefont {Chen}, \citenamefont {Cui},
  \citenamefont {Eriksen}, \citenamefont {Gao}, \citenamefont {Guo},
  \citenamefont {Hermann}, \citenamefont {Hermes}, \citenamefont {Koh},
  \citenamefont {Koval}, \citenamefont {Lehtola}, \citenamefont {Li},
  \citenamefont {Liu}, \citenamefont {Mardirossian}, \citenamefont {McClain},
  \citenamefont {Motta}, \citenamefont {Mussard}, \citenamefont {Pham},
  \citenamefont {Pulkin}, \citenamefont {Purwanto}, \citenamefont {Robinson},
  \citenamefont {Ronca}, \citenamefont {Sayfutyarova}, \citenamefont
  {Scheurer}, \citenamefont {Schurkus}, \citenamefont {Smith}, \citenamefont
  {Sun}, \citenamefont {Sun}, \citenamefont {Upadhyay}, \citenamefont {Wagner},
  \citenamefont {Wang}, \citenamefont {White}, \citenamefont {Whitfield},
  \citenamefont {Williamson}, \citenamefont {Wouters}, \citenamefont {Yang},
  \citenamefont {Yu}, \citenamefont {Zhu}, \citenamefont {Berkelbach},
  \citenamefont {Sharma}, \citenamefont {Sokolov},\ and\ \citenamefont
  {Chan}}]{Sun2020}%
  \BibitemOpen
  \bibfield  {author} {\bibinfo {author} {\bibfnamefont {Q.}~\bibnamefont
  {Sun}}, \bibinfo {author} {\bibfnamefont {X.}~\bibnamefont {Zhang}}, \bibinfo
  {author} {\bibfnamefont {S.}~\bibnamefont {Banerjee}}, \bibinfo {author}
  {\bibfnamefont {P.}~\bibnamefont {Bao}}, \bibinfo {author} {\bibfnamefont
  {M.}~\bibnamefont {Barbry}}, \bibinfo {author} {\bibfnamefont {N.~S.}\
  \bibnamefont {Blunt}}, \bibinfo {author} {\bibfnamefont {N.~A.}\ \bibnamefont
  {Bogdanov}}, \bibinfo {author} {\bibfnamefont {G.~H.}\ \bibnamefont {Booth}},
  \bibinfo {author} {\bibfnamefont {J.}~\bibnamefont {Chen}}, \bibinfo {author}
  {\bibfnamefont {Z.-H.}\ \bibnamefont {Cui}}, \bibinfo {author} {\bibfnamefont
  {J.~J.}\ \bibnamefont {Eriksen}}, \bibinfo {author} {\bibfnamefont
  {Y.}~\bibnamefont {Gao}}, \bibinfo {author} {\bibfnamefont {S.}~\bibnamefont
  {Guo}}, \bibinfo {author} {\bibfnamefont {J.}~\bibnamefont {Hermann}},
  \bibinfo {author} {\bibfnamefont {M.~R.}\ \bibnamefont {Hermes}}, \bibinfo
  {author} {\bibfnamefont {K.}~\bibnamefont {Koh}}, \bibinfo {author}
  {\bibfnamefont {P.}~\bibnamefont {Koval}}, \bibinfo {author} {\bibfnamefont
  {S.}~\bibnamefont {Lehtola}}, \bibinfo {author} {\bibfnamefont
  {Z.}~\bibnamefont {Li}}, \bibinfo {author} {\bibfnamefont {J.}~\bibnamefont
  {Liu}}, \bibinfo {author} {\bibfnamefont {N.}~\bibnamefont {Mardirossian}},
  \bibinfo {author} {\bibfnamefont {J.~D.}\ \bibnamefont {McClain}}, \bibinfo
  {author} {\bibfnamefont {M.}~\bibnamefont {Motta}}, \bibinfo {author}
  {\bibfnamefont {B.}~\bibnamefont {Mussard}}, \bibinfo {author} {\bibfnamefont
  {H.~Q.}\ \bibnamefont {Pham}}, \bibinfo {author} {\bibfnamefont
  {A.}~\bibnamefont {Pulkin}}, \bibinfo {author} {\bibfnamefont
  {W.}~\bibnamefont {Purwanto}}, \bibinfo {author} {\bibfnamefont {P.~J.}\
  \bibnamefont {Robinson}}, \bibinfo {author} {\bibfnamefont {E.}~\bibnamefont
  {Ronca}}, \bibinfo {author} {\bibfnamefont {E.~R.}\ \bibnamefont
  {Sayfutyarova}}, \bibinfo {author} {\bibfnamefont {M.}~\bibnamefont
  {Scheurer}}, \bibinfo {author} {\bibfnamefont {H.~F.}\ \bibnamefont
  {Schurkus}}, \bibinfo {author} {\bibfnamefont {J.~E.~T.}\ \bibnamefont
  {Smith}}, \bibinfo {author} {\bibfnamefont {C.}~\bibnamefont {Sun}}, \bibinfo
  {author} {\bibfnamefont {S.-N.}\ \bibnamefont {Sun}}, \bibinfo {author}
  {\bibfnamefont {S.}~\bibnamefont {Upadhyay}}, \bibinfo {author}
  {\bibfnamefont {L.~K.}\ \bibnamefont {Wagner}}, \bibinfo {author}
  {\bibfnamefont {X.}~\bibnamefont {Wang}}, \bibinfo {author} {\bibfnamefont
  {A.}~\bibnamefont {White}}, \bibinfo {author} {\bibfnamefont {J.~D.}\
  \bibnamefont {Whitfield}}, \bibinfo {author} {\bibfnamefont {M.~J.}\
  \bibnamefont {Williamson}}, \bibinfo {author} {\bibfnamefont
  {S.}~\bibnamefont {Wouters}}, \bibinfo {author} {\bibfnamefont
  {J.}~\bibnamefont {Yang}}, \bibinfo {author} {\bibfnamefont {J.~M.}\
  \bibnamefont {Yu}}, \bibinfo {author} {\bibfnamefont {T.}~\bibnamefont
  {Zhu}}, \bibinfo {author} {\bibfnamefont {T.~C.}\ \bibnamefont {Berkelbach}},
  \bibinfo {author} {\bibfnamefont {S.}~\bibnamefont {Sharma}}, \bibinfo
  {author} {\bibfnamefont {A.~Y.}\ \bibnamefont {Sokolov}}, \ and\ \bibinfo
  {author} {\bibfnamefont {G.~K.-L.}\ \bibnamefont {Chan}},\ }\href {\doibase
  10.1063/5.0006074} {\bibfield  {journal} {\bibinfo  {journal} {J. Chem.
  Phys.}\ }\textbf {\bibinfo {volume} {153}},\ \bibinfo {pages} {024109}
  (\bibinfo {year} {2020})}\BibitemShut {NoStop}%
\bibitem [{\citenamefont {Sch{\"a}fer}, \citenamefont {Huber},\ and\
  \citenamefont {Ahlrichs}(1994)}]{schafer1994fully}%
  \BibitemOpen
  \bibfield  {author} {\bibinfo {author} {\bibfnamefont {A.}~\bibnamefont
  {Sch{\"a}fer}}, \bibinfo {author} {\bibfnamefont {C.}~\bibnamefont {Huber}},
  \ and\ \bibinfo {author} {\bibfnamefont {R.}~\bibnamefont {Ahlrichs}},\
  }\href {\doibase 10.1063/1.463096} {\bibfield  {journal} {\bibinfo  {journal}
  {J. Chem. Phys.}\ }\textbf {\bibinfo {volume} {100}},\ \bibinfo {pages}
  {5829} (\bibinfo {year} {1994})}\BibitemShut {NoStop}%
\bibitem [{\citenamefont {Godbout}\ \emph {et~al.}(1992)\citenamefont
  {Godbout}, \citenamefont {Salahub}, \citenamefont {Andzelm},\ and\
  \citenamefont {Wimmer}}]{godbout1992optimization}%
  \BibitemOpen
  \bibfield  {author} {\bibinfo {author} {\bibfnamefont {N.}~\bibnamefont
  {Godbout}}, \bibinfo {author} {\bibfnamefont {D.~R.}\ \bibnamefont
  {Salahub}}, \bibinfo {author} {\bibfnamefont {J.}~\bibnamefont {Andzelm}}, \
  and\ \bibinfo {author} {\bibfnamefont {E.}~\bibnamefont {Wimmer}},\ }\href
  {\doibase 10.1139/v92-079} {\bibfield  {journal} {\bibinfo  {journal} {Can.
  J. Chem.}\ }\textbf {\bibinfo {volume} {70}},\ \bibinfo {pages} {560}
  (\bibinfo {year} {1992})}\BibitemShut {NoStop}%
\bibitem [{\citenamefont {Schreiber}\ \emph {et~al.}(2008)\citenamefont
  {Schreiber}, \citenamefont {Silva-Junior}, \citenamefont {Sauer},\ and\
  \citenamefont {Thiel}}]{schreiber2008benchmarks}%
  \BibitemOpen
  \bibfield  {author} {\bibinfo {author} {\bibfnamefont {M.}~\bibnamefont
  {Schreiber}}, \bibinfo {author} {\bibfnamefont {M.~R.}\ \bibnamefont
  {Silva-Junior}}, \bibinfo {author} {\bibfnamefont {S.~P.}\ \bibnamefont
  {Sauer}}, \ and\ \bibinfo {author} {\bibfnamefont {W.}~\bibnamefont
  {Thiel}},\ }\href {\doibase 10.1063/1.2889385} {\bibfield  {journal}
  {\bibinfo  {journal} {J. Chem. Phys.}\ }\textbf {\bibinfo {volume} {128}},\
  \bibinfo {pages} {134110} (\bibinfo {year} {2008})}\BibitemShut {NoStop}%
\bibitem [{\citenamefont {Silva-Junior}\ \emph {et~al.}(2008)\citenamefont
  {Silva-Junior}, \citenamefont {Schreiber}, \citenamefont {Sauer},\ and\
  \citenamefont {Thiel}}]{silva2008benchmarks}%
  \BibitemOpen
  \bibfield  {author} {\bibinfo {author} {\bibfnamefont {M.~R.}\ \bibnamefont
  {Silva-Junior}}, \bibinfo {author} {\bibfnamefont {M.}~\bibnamefont
  {Schreiber}}, \bibinfo {author} {\bibfnamefont {S.~P.}\ \bibnamefont
  {Sauer}}, \ and\ \bibinfo {author} {\bibfnamefont {W.}~\bibnamefont
  {Thiel}},\ }\href {\doibase 10.1063/1.2973541} {\bibfield  {journal}
  {\bibinfo  {journal} {J. Chem. Phys.}\ }\textbf {\bibinfo {volume} {129}},\
  \bibinfo {pages} {104103} (\bibinfo {year} {2008})}\BibitemShut {NoStop}%
\bibitem [{\citenamefont {van Setten}\ \emph
  {et~al.}(2015{\natexlab{b}})\citenamefont {van Setten}, \citenamefont
  {Caruso}, \citenamefont {Sharifzadeh}, \citenamefont {Ren}, \citenamefont
  {Scheffler}, \citenamefont {Liu}, \citenamefont {Lischner}, \citenamefont
  {Lin}, \citenamefont {Deslippe}, \citenamefont {Louie} \emph
  {et~al.}}]{van2015gw}%
  \BibitemOpen
  \bibfield  {author} {\bibinfo {author} {\bibfnamefont {M.~J.}\ \bibnamefont
  {van Setten}}, \bibinfo {author} {\bibfnamefont {F.}~\bibnamefont {Caruso}},
  \bibinfo {author} {\bibfnamefont {S.}~\bibnamefont {Sharifzadeh}}, \bibinfo
  {author} {\bibfnamefont {X.}~\bibnamefont {Ren}}, \bibinfo {author}
  {\bibfnamefont {M.}~\bibnamefont {Scheffler}}, \bibinfo {author}
  {\bibfnamefont {F.}~\bibnamefont {Liu}}, \bibinfo {author} {\bibfnamefont
  {J.}~\bibnamefont {Lischner}}, \bibinfo {author} {\bibfnamefont
  {L.}~\bibnamefont {Lin}}, \bibinfo {author} {\bibfnamefont {J.~R.}\
  \bibnamefont {Deslippe}}, \bibinfo {author} {\bibfnamefont {S.~G.}\
  \bibnamefont {Louie}},  \emph {et~al.},\ }\href {\doibase
  10.1021/acs.jctc.5b00453} {\bibfield  {journal} {\bibinfo  {journal} {J.
  Chem. Theory Comput.}\ }\textbf {\bibinfo {volume} {11}},\ \bibinfo {pages}
  {5665} (\bibinfo {year} {2015}{\natexlab{b}})}\BibitemShut {NoStop}%
\bibitem [{pub(2021)}]{pubchem}%
  \BibitemOpen
  \href@noop {} {\enquote {\bibinfo {title} {National center for biotechnology
  information},}\ } (\bibinfo {year} {2021}),\ \bibinfo {note} {pubChem
  Compound Summary for CID 71353974 and 101946798.
  https://pubchem.ncbi.nlm.nih.gov. Accessed July 1, 2021.}\BibitemShut {Stop}%
\bibitem [{\citenamefont {Wilson}, \citenamefont {McKenzie-Sell},\ and\
  \citenamefont {Barnard}(2014)}]{wilson2014shape}%
  \BibitemOpen
  \bibfield  {author} {\bibinfo {author} {\bibfnamefont {H.~F.}\ \bibnamefont
  {Wilson}}, \bibinfo {author} {\bibfnamefont {L.}~\bibnamefont
  {McKenzie-Sell}}, \ and\ \bibinfo {author} {\bibfnamefont {A.~S.}\
  \bibnamefont {Barnard}},\ }\href {\doibase 10.1039/C4TC01312C} {\bibfield
  {journal} {\bibinfo  {journal} {J. Mater. Chem. C}\ }\textbf {\bibinfo
  {volume} {2}},\ \bibinfo {pages} {9451} (\bibinfo {year} {2014})}\BibitemShut
  {NoStop}%
\bibitem [{\citenamefont {Wolkin}\ \emph {et~al.}(1999)\citenamefont {Wolkin},
  \citenamefont {Jorne}, \citenamefont {Fauchet}, \citenamefont {Allan},\ and\
  \citenamefont {Delerue}}]{wolkin1999electronic}%
  \BibitemOpen
  \bibfield  {author} {\bibinfo {author} {\bibfnamefont {M.}~\bibnamefont
  {Wolkin}}, \bibinfo {author} {\bibfnamefont {J.}~\bibnamefont {Jorne}},
  \bibinfo {author} {\bibfnamefont {P.}~\bibnamefont {Fauchet}}, \bibinfo
  {author} {\bibfnamefont {G.}~\bibnamefont {Allan}}, \ and\ \bibinfo {author}
  {\bibfnamefont {C.}~\bibnamefont {Delerue}},\ }\href {\doibase
  10.1103/PhysRevLett.82.197} {\bibfield  {journal} {\bibinfo  {journal} {Phys.
  Rev. Lett.}\ }\textbf {\bibinfo {volume} {82}},\ \bibinfo {pages} {197}
  (\bibinfo {year} {1999})}\BibitemShut {NoStop}%
\bibitem [{\citenamefont {Huang}\ \emph {et~al.}(2016)\citenamefont {Huang},
  \citenamefont {Harris}, \citenamefont {Krzyaniak}, \citenamefont {Margulies},
  \citenamefont {Dyar}, \citenamefont {Lindquist}, \citenamefont {Wu},
  \citenamefont {Roznyatovskiy}, \citenamefont {Wu}, \citenamefont {Young}
  \emph {et~al.}}]{huang2016photoinduced}%
  \BibitemOpen
  \bibfield  {author} {\bibinfo {author} {\bibfnamefont {G.-J.}\ \bibnamefont
  {Huang}}, \bibinfo {author} {\bibfnamefont {M.~A.}\ \bibnamefont {Harris}},
  \bibinfo {author} {\bibfnamefont {M.~D.}\ \bibnamefont {Krzyaniak}}, \bibinfo
  {author} {\bibfnamefont {E.~A.}\ \bibnamefont {Margulies}}, \bibinfo {author}
  {\bibfnamefont {S.~M.}\ \bibnamefont {Dyar}}, \bibinfo {author}
  {\bibfnamefont {R.~J.}\ \bibnamefont {Lindquist}}, \bibinfo {author}
  {\bibfnamefont {Y.}~\bibnamefont {Wu}}, \bibinfo {author} {\bibfnamefont
  {V.~V.}\ \bibnamefont {Roznyatovskiy}}, \bibinfo {author} {\bibfnamefont
  {Y.-L.}\ \bibnamefont {Wu}}, \bibinfo {author} {\bibfnamefont {R.~M.}\
  \bibnamefont {Young}},  \emph {et~al.},\ }\href {\doibase
  10.1021/acs.jpcb.5b10806} {\bibfield  {journal} {\bibinfo  {journal} {J.
  Phys. Chem. B}\ }\textbf {\bibinfo {volume} {120}},\ \bibinfo {pages} {756}
  (\bibinfo {year} {2016})}\BibitemShut {NoStop}%
\bibitem [{\citenamefont {Martin}(2003)}]{Martin2003}%
  \BibitemOpen
  \bibfield  {author} {\bibinfo {author} {\bibfnamefont {R.~L.}\ \bibnamefont
  {Martin}},\ }\href {\doibase 10.1063/1.1558471} {\bibfield  {journal}
  {\bibinfo  {journal} {J. Chem. Phys.}\ }\textbf {\bibinfo {volume} {118}},\
  \bibinfo {pages} {4775} (\bibinfo {year} {2003})}\BibitemShut {NoStop}%
\bibitem [{\citenamefont {Shishkin}, \citenamefont {Marsman},\ and\
  \citenamefont {Kresse}(2007)}]{Shishkin2007}%
  \BibitemOpen
  \bibfield  {author} {\bibinfo {author} {\bibfnamefont {M.}~\bibnamefont
  {Shishkin}}, \bibinfo {author} {\bibfnamefont {M.}~\bibnamefont {Marsman}}, \
  and\ \bibinfo {author} {\bibfnamefont {G.}~\bibnamefont {Kresse}},\ }\href
  {\doibase 10.1103/physrevlett.99.246403} {\bibfield  {journal} {\bibinfo
  {journal} {Phys. Rev. Lett.}\ }\textbf {\bibinfo {volume} {99}},\ \bibinfo
  {pages} {246403} (\bibinfo {year} {2007})}\BibitemShut {NoStop}%
\bibitem [{\citenamefont {Romaniello}, \citenamefont {Guyot},\ and\
  \citenamefont {Reining}(2009)}]{Romaniello2009}%
  \BibitemOpen
  \bibfield  {author} {\bibinfo {author} {\bibfnamefont {P.}~\bibnamefont
  {Romaniello}}, \bibinfo {author} {\bibfnamefont {S.}~\bibnamefont {Guyot}}, \
  and\ \bibinfo {author} {\bibfnamefont {L.}~\bibnamefont {Reining}},\ }\href
  {\doibase 10.1063/1.3249965} {\bibfield  {journal} {\bibinfo  {journal} {J.
  Chem. Phys.}\ }\textbf {\bibinfo {volume} {131}},\ \bibinfo {pages} {154111}
  (\bibinfo {year} {2009})}\BibitemShut {NoStop}%
\bibitem [{\citenamefont {Maggio}\ and\ \citenamefont
  {Kresse}(2017)}]{Maggio2017}%
  \BibitemOpen
  \bibfield  {author} {\bibinfo {author} {\bibfnamefont {E.}~\bibnamefont
  {Maggio}}\ and\ \bibinfo {author} {\bibfnamefont {G.}~\bibnamefont
  {Kresse}},\ }\href {\doibase 10.1021/acs.jctc.7b00586} {\bibfield  {journal}
  {\bibinfo  {journal} {J. Chem. Theory Comput.}\ }\textbf {\bibinfo {volume}
  {13}},\ \bibinfo {pages} {4765} (\bibinfo {year} {2017})}\BibitemShut
  {NoStop}%
\bibitem [{\citenamefont {Lewis}\ and\ \citenamefont
  {Berkelbach}(2019)}]{Lewis2019}%
  \BibitemOpen
  \bibfield  {author} {\bibinfo {author} {\bibfnamefont {A.~M.}\ \bibnamefont
  {Lewis}}\ and\ \bibinfo {author} {\bibfnamefont {T.~C.}\ \bibnamefont
  {Berkelbach}},\ }\href {\doibase 10.1021/acs.jctc.8b00995} {\bibfield
  {journal} {\bibinfo  {journal} {J. Chem. Theory Comput.}\ }\textbf {\bibinfo
  {volume} {15}},\ \bibinfo {pages} {2925} (\bibinfo {year}
  {2019})}\BibitemShut {NoStop}%
\bibitem [{\citenamefont {Biermann}, \citenamefont {Aryasetiawan},\ and\
  \citenamefont {Georges}(2003)}]{Biermann2003}%
  \BibitemOpen
  \bibfield  {author} {\bibinfo {author} {\bibfnamefont {S.}~\bibnamefont
  {Biermann}}, \bibinfo {author} {\bibfnamefont {F.}~\bibnamefont
  {Aryasetiawan}}, \ and\ \bibinfo {author} {\bibfnamefont {A.}~\bibnamefont
  {Georges}},\ }\href {\doibase 10.1103/physrevlett.90.086402} {\bibfield
  {journal} {\bibinfo  {journal} {Phys. Rev. Lett.}\ }\textbf {\bibinfo
  {volume} {90}},\ \bibinfo {pages} {086402} (\bibinfo {year}
  {2003})}\BibitemShut {NoStop}%
\bibitem [{\citenamefont {Zhu}\ and\ \citenamefont
  {Chan}(2021{\natexlab{b}})}]{Zhu2021a}%
  \BibitemOpen
  \bibfield  {author} {\bibinfo {author} {\bibfnamefont {T.}~\bibnamefont
  {Zhu}}\ and\ \bibinfo {author} {\bibfnamefont {G.~K.-L.}\ \bibnamefont
  {Chan}},\ }\href {\doibase 10.1103/physrevx.11.021006} {\bibfield  {journal}
  {\bibinfo  {journal} {Phys. Rev. X}\ }\textbf {\bibinfo {volume} {11}},\
  \bibinfo {pages} {021006} (\bibinfo {year} {2021}{\natexlab{b}})}\BibitemShut
  {NoStop}%
\end{thebibliography}
\end{document}